\documentclass[11pt,a4paper]{article}
\pdfoutput=1
\usepackage{jheppub}

\usepackage[numbers,sort&compress]{natbib}
\usepackage{subcaption}

\renewcommand*{\le}{\left}
\newcommand*{\ri}{\right}

\renewcommand*{\a}{\alpha}
\renewcommand*{\b}{\beta}
\newcommand*{\g}{\gamma}
\renewcommand*{\d}{\delta}
\newcommand*{\e}{\epsilon}
\newcommand*{\h}{\eta}
\newcommand*{\q}{\theta}

\newcommand*{\m}{\mu}
\newcommand*{\n}{\nu}
\renewcommand*{\r}{\rho}
\newcommand*{\s}{\sigma}
\renewcommand*{\t}{\tau}
\newcommand*{\f}{\phi}
\renewcommand*{\c}{\chi}
\newcommand*{\F}{\Phi}

\newcommand*{\tfive}{T_\mathrm{M5}}
\newcommand*{\ttwo}{T_\mathrm{M2}}

\newcommand*{\see}{S_\mathrm{EE}}

\newcommand*{\p}{\partial}
\newcommand*{\diff}{\mathrm{d}} % Upright d for derivatives
\newcommand*{\ct}{\tilde{c}}

\usepackage{xcolor}

\title{Holographic entanglement entropy from probe M-theory branes}
\author{Ronnie Rodgers}
\affiliation{STAG Research Centre, Physics and Astronomy, University of Southampton,\\
	Highfield, Southampton SO17 1BJ, United Kingdom}
\emailAdd{r.j.rodgers@soton.ac.uk}
\abstract{We compute the holographic entanglement entropy contribution from planar two-dimensional defects in six-dimensional \(\mathcal{N}=(2,0)\) superconformal field theory, holographically dual to probe M2- and M5-branes in \(AdS_7 \times S^4\).  In particular, we test the viability of the universal contribution of the defect to entanglement entropy as a candidate \(C\)-function. We find that this coefficient is not monotonic under defect renormalization group flows triggered by the vacuum expectation value of a marginal operator. Another candidate \(C\)-function, the on-shell action inside the entanglement wedge, monotonically decreases under the flows we study.
}

\begin{document}
%%%%%%%%%%%%%%%%%%%%%%%%%%%%%%%%%%%%%%%%%%%%%%%%%%
%%%%%%%%%%%%%%%%%%%%%%%%%%%%%%%%%%%%%%%%%%%%%%%%%%

\maketitle

%%%%%%%%%%%%%%%%%%%%%%%%%%%%%%%%%%%%%%%%%%%%%%%%%%
%%%%%%%%%%%%%%%%%%%%%%%%%%%%%%%%%%%%%%%%%%%%%%%%%%
\section{Introduction}
%%%%%%%%%%%%%%%%%%%%%%%%%%%%%%%%%%%%%%%%%%%%%%%%%%
%%%%%%%%%%%%%%%%%%%%%%%%%%%%%%%%%%%%%%%%%%%%%%%%%%

%%%%%%%%%%%%%%%%%%%%%%%%%%%%%%%%%%%%%%%%%%%%%%%%%%
%%%%%%%%%%%%%%%%%%%%%%%%%%%%%%%%%%%%%%%%%%%%%%%%%%
\subsection{Background and motivation}
%%%%%%%%%%%%%%%%%%%%%%%%%%%%%%%%%%%%%%%%%%%%%%%%%%
%%%%%%%%%%%%%%%%%%%%%%%%%%%%%%%%%%%%%%%%%%%%%%%%%%

M-theory plays an important role in string theory. The different superstring theories, arising as limits of M-theory under different compactifications, are united by a web of dualities which are interpreted as symmetries of M-theory. Despite much progress, many fundamental questions about M-theory remain open.

The low energy limit of M-theory is believed to be eleven-dimensional supergravity (11D SUGRA), with field content consisting of the metric, a gravitino, and a three-form gauge potential. The three-form naturally couples electrically to three-dimensional objects, called M2-branes, and magnetically to six-dimensional objects, called M5-branes. These branes are believed to be fundamental objects of M-theory.

In string theory, the low energy excitations of a stack of coincident D-branes are described by supersymmetric Yang-Mills (SYM) theory \cite{Witten:1995im}. Similarly, in M-theory the low energy excitations of a stack of M2-branes are described by Aharony-Bergman-Jafferis-Maldacena (ABJM) theory, maximally supersymmetric Chern-Simons theory coupled to matter \cite{Bagger:2006sk,Gustavsson:2007vu,Bagger:2007jr,Bagger:2007vi,Aharony:2008ug}.

The theory describing the low energy excitations of a stack of \(M > 1\) M5-branes is not known, but some information may be obtained from supergravity. In 11D SUGRA, a stack of flat M5-branes corresponds to a certain solitonic solution of the equations of motion. This solution preserves \(\mathcal{N}=(2,0)\) supersymmetry (SUSY) with \(SO(5)\) R-symmetry corresponding to rotations in the five directions normal to the branes. The world volume fields form \(M\) copies of the tensor multiplet of \(\mathcal{N} = (2,0)\) SUSY, which together are expected to realize a gauge multiplet for an \(\mathfrak{su}(M)\) gauge algebra \cite{Strominger:1995ac}. Various supergavity calculations indicate that the number of massless degrees of freedom in the theory scales as \(M^3\) at large \(M\) \cite{Klebanov:1996un,Freed:1998tg,Henningson:1998gx,Harvey:1998bx}.

Holography provides a powerful tool to study the M5-brane theory. When the number of branes \(M\) in the stack is much larger than one, the world volume theory is expected to be holographically dual to 11D SUGRA on \(AdS_7 \times S^4\) \cite{Maldacena:1997re}; in particular the theory is expected to be a superconformal field theory (SCFT).

Beyond the important role that the \(\mathcal{N}=(2,0)\) theory plays in M-theory, it is notable among quantum field theories as it is the maximally supersymmetric theory in six-dimensions, which is the largest number of dimensions in which superconformal symmetry is possible~\cite{Nahm:1977tg}. Study of compactifications of this theory and other 6D SCFTs has revealed intriguing properties of lower dimensional quantum field theories, such as the conjectured relation between four-dimensional \(\mathcal{N}=2\) gauge theories and Liouville or Toda theories in two dimensions \cite{Alday:2009aq,Wyllard:2009hg}.

%%%%%%%%%%%%%%%%%%%%%%%%%%%%%%%%%%%%%%%%%%%%%%%%%%
%%%%%%%%%%%%%%%%%%%%%%%%%%%%%%%%%%%%%%%%%%%%%%%%%%
\subsection{A defect central charge from entanglement entropy}
%%%%%%%%%%%%%%%%%%%%%%%%%%%%%%%%%%%%%%%%%%%%%%%%%%
%%%%%%%%%%%%%%%%%%%%%%%%%%%%%%%%%%%%%%%%%%%%%%%%%%

We will use holography to study (1+1)-dimensional defects in \(\mathcal{N} = (2,0)\)~superconformal field theory (SCFT), dual to probe branes in \(AdS_7 \times S^4\) which touch the boundary of \(AdS\). In particular, we will compute the contribution of the defect to entanglement entropy.

A subset of the defects we study are expected to be one-half BPS Wilson surface operators~\cite{Lunin:2007ab,Chen:2007ir,Mori:2014tca}, and in these cases our results reproduce the probe limit of the calculations in refs.~\cite{Gentle:2015jma, us_entanglement}. We will also study solutions dual to defect renormalization group (RG) flows between these Wilson surfaces and conformal defects dual to bundles of M2-branes.

\begin{figure}
	\begin{center}
	\includegraphics{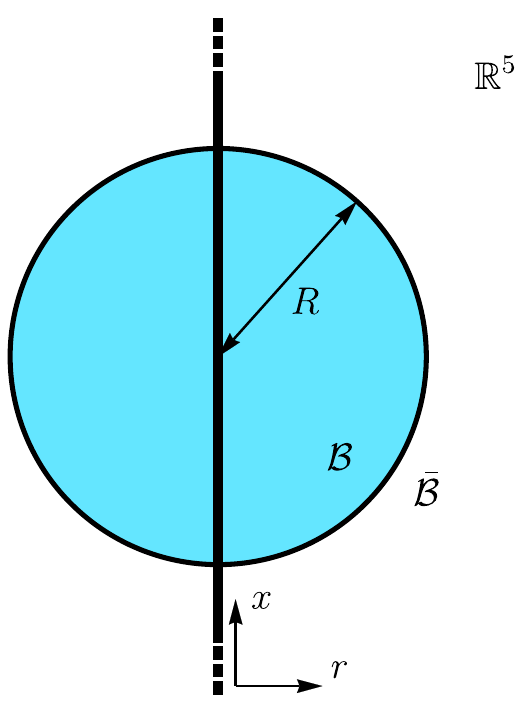}
	\caption{We compute the entanglement entropy between a spatial subregion~\(\mathcal{B}\), and its complement~\(\bar{\mathcal{B}}\). Throughout, we will take \(\mathcal{B}\) to be a five-dimensional ball of radius \(R\) (shaded blue in the diagram), centered on a planar (1+1)-dimensional defect (the thick, vertical line, with dots indicating that the defect extends to infinity). The intersection of \(\mathcal{B}\) with the defect is an interval of length \(2R\), so it is not surprising that the contribution of the defect to the entanglement entropy takes the form~\eqref{eq:defect_ee} of the entanglement entropy for a single interval in a 2D CFT. The coordinates \(x\) and \(r\) are defined in section~\ref{sec:actions}.}
	\label{fig:entangling_region}
	\end{center}
\end{figure}
We wish to compute the entanglement entropy of a spatial subregion with its complement. For simplicity, we will restrict to planar defects, and choose the entangling surface to be a sphere of radius \(R\) centered at a point on the defect, as illustrated in figure~\ref{fig:entangling_region}. In a CFT in six dimensions, the entanglement entropy \(\see\) of a spherical subregion takes the form~\cite{Ryu:2006bv,Ryu:2006ef}
\begin{equation} \label{eq:entanglement_general_form}
	\see = p_4 \frac{R^4}{\e^4} + p_2 \frac{R^2}{\e^2} + p_L \log\le( \frac{2R}{\e} \ri) + \mathcal{O}(\e^0),
\end{equation}
where \(\e\) is an ultraviolet cutoff. The coefficients \(p_2\) and \(p_4\) are scheme dependent --- they are not invariant under multiplicative changes in the cutoff --- while the coefficient \(p_L\) of the logarithm is scheme independent. For the vacuum of the \(\mathcal{N} = (2,0)\) theory, holographically dual to 11D SUGRA on \(AdS_7 \times S^4\), it is given by~\cite{Ryu:2006bv}
\begin{equation} \label{eq:m5_entanglement_coefficient}
	p_L = \frac{4}{3} M^3.
\end{equation}

The presence of a two-dimensional planar defect will modify the coefficient of the logarithmic term in equation~\eqref{eq:entanglement_general_form}. Defining \(\see^{(0)}\) as the entanglement entropy without the defect, i.e. \eqref{eq:entanglement_general_form} with in particular the coefficient of the logarithm given by \eqref{eq:m5_entanglement_coefficient}, we define the contribution to entanglement entropy from the defect as\footnote{We assume that the same regularization prescription is used for the entanglement entropy in the defect CFT as in the theory without the defect.}
\begin{equation} \label{eq:defect_ee}
	\see^{(1)} \equiv \see - \see^{(0)} = \frac{b}{3} \log \le( \frac{2R}{\e} \ri) + \mathcal{O}(\e^0),
\end{equation}
with a coefficient \(b\) to be determined.

The entanglement entropy of an interval of length \(\propto R\) in a two-dimensional CFT takes the same form, with \(b\) replaced by the central charge \(c\) of the CFT~\cite{Calabrese:2004eu}. By analogy, we will refer to the coefficient \(b\) in \eqref{eq:defect_ee} as the central charge of the defect. The central charge of a 2D CFT measures the number of degrees of freedom in the theory. One of the questions we will seek to address is whether \(b\) similarly measures degrees of freedom on the defect.

For defect RG flows we may naturally define an \(R\) dependent \(b\)-function,
\begin{equation} \label{eq:b_function}
    b(R) \equiv 3 R \frac{ \diff \see^{(1)} }{ \diff R }.
\end{equation}
In a CFT \(b(R)\) is a constant, equal to the central charge \(b\).\footnote{Note that in a CFT the \(\mathcal{O}(\e^0)\) term in \eqref{eq:defect_ee} must be independent of \(R\), since the entanglement entropy is dimensionless and there are no other scales which may be combined with \(R\) to yield a dimensionless quantity.} Along an RG flow, the \(b\)-function interpolates between the putative defect central charges of the fixed points,
\begin{equation}
    \lim_{R\to0} b(R) = b_\mathrm{UV},
    \quad
    \lim_{R\to\infty} b(R) =  b_{\mathrm{IR}},
\end{equation}
where \(b_\mathrm{UV}\) and \(b_\mathrm{IR}\) are the central charges of the ultraviolet (UV) and infrared (IR) fixed points, respectively.

In two-dimensions, the logarithmic derivative of the single-interval entanglement entropy with respect to the length of the interval is a \(C\)-function. This means that it satisfies a (strong) monotonicity theorem \cite{Casini:2006es}, so it decreases monotonically along any RG flow. On the other hand, we find that for the flows we study \(b(R)\) is not monotonic, and may be larger in the IR than in the UV. This provides an obstruction to interpreting \(b\) as a measure of degrees of freedom.

It has recently been argued \cite{Kobayashi:2018lil} that in general defect CFTs the entanglement entropy does not provide a good candidate to satisfy a monotonicity theorem. Instead, ref.~\cite{Kobayashi:2018lil} proposed that the defect contribution to the free energy, either on a sphere or in hyperbolic space, should increase monotonically along RG flows.\footnote{This is a generalization of the \(F\)-theorem for theories in three dimensions without a defect~\cite{Jafferis:2011zi,Klebanov:2011gs}.} In particular, the free energy should be larger in the UV than in the IR. We find that this is indeed the case for the free energy in hyperbolic space in the flows that we study. 

%%%%%%%%%%%%%%%%%%%%%%%%%%%%%%%%%%%%%%%%%%%%%%%%%%
%%%%%%%%%%%%%%%%%%%%%%%%%%%%%%%%%%%%%%%%%%%%%%%%%%
\subsection{Outline}
%%%%%%%%%%%%%%%%%%%%%%%%%%%%%%%%%%%%%%%%%%%%%%%%%%
%%%%%%%%%%%%%%%%%%%%%%%%%%%%%%%%%%%%%%%%%%%%%%%%%%

This paper is organised as follows. In section \ref{sec:actions} we review the actions describing single M2- and M5-branes, and the holographic calculation of entanglement entropy in the presence of probe branes. Section \ref{sec:m2} describes the entanglement entropy contribution from a single M2-brane, which provides a simple first example and allows us to establish some notation. In sections \ref{sec:antisymmetric} and \ref{sec:symmetric} we compute the contribution to entanglement entropy from M5-branes wrapping an \(S^3\) internal to the \(S^4\) or \(AdS_7\) factors of the background geometry, respectively. We also examine possible \(C\)-functions provided by the on-shell action. We close with some discussion of the results and outlook for the future in section \ref{sec:discussion}.

This paper is a companion to~\cite{us_entanglement}, which studies the holographic entanglement entropy of arbitrary representation Wilson surfaces, dual to solutions of 11D SUGRA describing back-reacted brane configurations. In addition, all of the M5-brane embeddings discussed in this paper are analogous to certain D3- and D5-brane embeddings in \(AdS_5 \times S^5\), which have been studied in detail in \cite{Kumar:2016jxy,Kumar:2017vjv}. Our analysis will closely follow that of ref.~\cite{Kumar:2017vjv}, in which the entanglement entropy of these D-brane solutions was computed.

%%%%%%%%%%%%%%%%%%%%%%%%%%%%%%%%%%%%%%%%%%%%%%%%%%
%%%%%%%%%%%%%%%%%%%%%%%%%%%%%%%%%%%%%%%%%%%%%%%%%%
\section{Probe branes in \(AdS_7 \times S^4\)}
\label{sec:actions}
%%%%%%%%%%%%%%%%%%%%%%%%%%%%%%%%%%%%%%%%%%%%%%%%%%
%%%%%%%%%%%%%%%%%%%%%%%%%%%%%%%%%%%%%%%%%%%%%%%%%%

The bosonic fields of 11D SUGRA are the metric \(G\) and three-form gauge potential \(C_3\), with field strength \(F_4 = \diff C_3\). The equations of motion for these fields admit solutions corresponding to a flat stack of \(M\) M5-branes,
\begin{subequations}
\begin{align}
	\label{eq:metric_with_h}
	\diff s^2 &= G_{MN} \diff x^M \diff x^N = h(\r)^{-1/3} \h_{\m\n} \diff x^\m \diff x^\n
	+ h(\r)^{2/3}\le(\diff \r^2 + \r^2 \diff s_{S^4}^2 \ri),
	\\
	F_4 &= \r^4 h'(\r) \diff s_{S^4},
\end{align}
\end{subequations}
where \(h(\r) = 1 + L^3/\r^3\), \(\h_{\m\n}\) is the 6-dimensional Minkowski metric, and \(\diff s_{S^4}\) is the volume form of a unit round \(S^4\), which we parameterise with a polar coordinate \(\q\) and three azimuthal coordinates coordinates \(\chi_{1,2,3}\). The parameter \(L\) is related to the Planck length \(\ell_P\) and the number of M5-branes by \(L = (\pi M)^{1/3} \ell_P\). Unless otherwise specified we will take the near-horizon limit \(\r \ll L\). In this limit the metric \eqref{eq:metric_with_h} becomes that of \(AdS_7 \times S^4\), with the radii of \(S^4\) and \(AdS_7\) equal to \(L\) and \(2L\), respectively. 11D~SUGRA on this background is expected to be holographically dual to the world volume theory describing the stack of M5-branes.

This work deals with (1+1)-dimensional planar defects, which we will take to span the \((x^0,x^1)\) plane. Let us define \(t \equiv x^0\) and \(x \equiv x^1\), and work with a spherical coordinate system in the remaining transverse directions on the boundary, with radial coordinate \(r^2 = \sum_{i=2}^5 (x^i)^2\) and three angular coordinates \(\f_{1,2,3}\) parameterizing an \(S^3\). It will also be convenient to define a new holographic radial coordinate, \(z\), by \(\r = 4 L^3/z^2\). After these coordinate transformations, the metric and gauge field strength in the near-horizon limit become
\begin{subequations}\label{eq:ads_solution}
\begin{align}
    \diff s^2 &= \frac{4L^2}{z^2} \le(-\diff t^2 + \diff x^2 + \diff r^2 + r^2  \diff s_{S^3}^2 + \diff z^2\ri) + L^2 \diff s_{S^4}^2 ,
    \label{eq:ads_metric}
	\\
	F_4 &= - 3 L^3 \diff s_{S^4},
	\label{eq:ads_field_strength}
\end{align}
\end{subequations}
with the boundary of \(AdS\) at \(z=0\). The metric factor \(\diff s_{S^3}^2\) is the metric on the round \(S^3\) parameterized by \(\f_{1,2,3}\). We will choose a gauge in which
\begin{subequations}
\begin{align}
	C_3 &= L^3 \le(3 \cos \q - \cos^3 \q - 2\ri) \sin^2\c_1 \sin\c_2 \, \diff \c_1 \wedge \diff \c_2 \wedge \diff \c_3,
	\label{eq:c3}
	\\
	C_6 &= \le( \frac{2 L}{z} \ri)^6 r^3 \sin^2 \f_1 \sin \f_2 \, \diff t \wedge \diff x \wedge \diff r \wedge \diff \f_1 \wedge \diff \f_2 \wedge \diff \f_3.
	\label{eq:flat_slicing_c6}
\end{align}
\end{subequations}
In particular \(C_3\) vanishes at the north pole of the \(S^4\), \(\q = 0\), so as to match with the calculations in ref.~\cite{Camino:2001at}.

%%%%%%%%%%%%%%%%%%%%%%%%%%%%%%%%%%%%%%%%%%%%%%%%%%
%%%%%%%%%%%%%%%%%%%%%%%%%%%%%%%%%%%%%%%%%%%%%%%%%%
\subsection{M-brane actions}
%%%%%%%%%%%%%%%%%%%%%%%%%%%%%%%%%%%%%%%%%%%%%%%%%%
%%%%%%%%%%%%%%%%%%%%%%%%%%%%%%%%%%%%%%%%%%%%%%%%%%

In the presence of an M2- or M5-brane, the action for 11D SUGRA becomes
\begin{equation}
	S = S_{11} + S_\mathrm{brane},
\end{equation}
where \(S_{11}\) is the bulk action for the eleven-dimensional supergravity fields, and \(S_\mathrm{brane}  = S_\mathrm{M2}\) or \(S_\mathrm{M5}\) is a contribution localized to the brane. We will always work in the probe limit, in which it is a good approximation to neglect the back-reaction of the brane on the metric and gauge field, which we may therefore take to be the \(AdS_7 \times S^4\) solution \eqref{eq:ads_solution}. The brane action \(S_\mathrm{brane}\) is then an action for the world volume fields of the brane, decoupled from the bulk supergravity fields.\footnote{See~\cite{Karch:2015kfa,Robinson:2017sup} for recent evidence for the validity of the probe approximation in holography.}

For a single M2-brane, the bosonic world volume fields are eight real scalars, which determine the embedding of the brane. They are described by the action \cite{Bergshoeff:1987cm}
\begin{equation} \label{eq:m2_action}
	S_\mathrm{M2} = - \ttwo \int_\Sigma \diff^3 \xi \sqrt{-\det g} + \ttwo \int_\Sigma P[C_3],
\end{equation}
where \(P\) denotes the pullback onto the brane of a bulk supergravity field, \(g \equiv P[G]\), and \(\xi\) are coordinates on the brane world volume \(\Sigma\). The tension \(\ttwo\) is related to the Planck length by \(\ttwo = 1/4\pi^2\ell_P^3\).

For a single M5-brane, the bosonic fields are five real scalar fields and an abelian two-form gauge field \(A\), with self-dual field strength \(F_3 \equiv \diff A\). Various formulations of the action for an M5-brane exist, which impose the self-duality constraint in different ways ~\cite{Pasti:1997gx,Bandos:1997ui,Aganagic:1997zq,Townsend:1995af,Schwarz:1997mc, Ko:2013dka}. The different formulations are believed to be equivalent, at least classically \cite{Bandos:1997gm,Ko:2013dka}. We will use the approach of  Pasti, Sorokin and Tonin (PST) \cite{Pasti:1997gx,Bandos:1997ui,Aganagic:1997zq}, which we find to be the simplest for our purposes. In this approach the self-duality constraint is imposed by an additional local symmetry due to the presence of an auxiliary scalar field \(a\).

The bosonic part of the PST action for a single M5-brane is \cite{Pasti:1997gx}
\begin{align} \label{eq:pst_action}
	S_\mathrm{M5} &= - \tfive \int_\Sigma d^6 \xi \le[
		\sqrt{-\det\le(g + i \tilde H\ri)} + \frac{\sqrt{-\det g}}{4 (\p a)^2} \p_m a H^{*mnl} H_{mnp} \p^p a
	\ri]
	\nonumber \\ & \hspace{4cm}
	+ \tfive \int_\Sigma \le(P[C_6] + \frac{1}{2} F_3 \wedge P[C_3] \ri).
\end{align}
where \(H \equiv F_3 + P[C_3]\), \(H^{*mnl} \equiv \frac{1}{6\sqrt{-g}}\e^{mnlpqr} H_{pqr}\), and \(\tilde H_{mn} \equiv H^{*}_{mn}{}^l \p_l a /\sqrt{(\p a)^2}\). The tension is given in terms of the Planck length by \(\tfive = 1/(2\pi)^5 \ell_P^6\).

We seek brane embeddings that span the defect (the (\(t,x\)) plane) at the boundary. Near the boundary, the geometry of the brane's world volume will be \(AdS_3 \times S^3\), where the \(S^3\) is either the \(S^3\) inside \(AdS_7\) parameterised by the \(\f_i\), or is internal to the \(S^4\), parameterised by the \(\chi_i\). This \(S^3\) is supported by flux of the world volume gauge field \(A\), sourced by M2-brane charge dissolved within the M5-brane. The total number of dissolved M2-branes \(N\) is given by the flux quantization condition \cite{Camino:2001at}
\begin{equation} \label{eq:flux_quantization}
	N = \frac{T_\mathrm{M2}}{2\pi} \int_{S^3} F_3.
\end{equation}

A subset of the M5-brane embeddings we consider are believed to be dual to half-BPS Wilson surface operators, in representations described by Young tableaux with number of boxes \(N\) of the order of the rank (\(\sim M\)) of the gauge algebra \cite{Lunin:2007ab,Chen:2007ir,Mori:2014tca}. This is analogous to the holographic description of Wilson lines in \(\mathcal{N}=4\) super Yang-Mills theory (SYM) by D-branes \cite{Drukker:1999zq,Drukker:2005kx,Gomis:2006sb,Gomis:2006im}. The restriction that \(N\) is at most of order \(M\) comes from the probe approximation, and applies to all of the solutions we study.

To holographically describe Wilson lines in SYM, one must add boundary terms to the D-brane action which implement a Legendre transformation with respect to the brane's position and gauge field \cite{Drukker:1999zq,Drukker:2005kx}. The former is needed because a string describing a Wilson line obeys complementary boundary conditions to a string ending on a D-brane. The latter fixes the total amount of fundamental string charge dissolved in the brane, and thus the representation of the Wilson line.

For M2- and M5-branes, we will use an analogous boundary term,
\begin{equation} \label{eq:boundary_term}
    S_\mathrm{bdy} = -\int_{\p\Sigma} d^p\s \, \r \frac{\d S_\mathrm{brane}}{\d(\p_n \r)} = \frac{1}{2} \int_{\p\Sigma} d^p\s \, z \frac{\d S_\mathrm{brane}}{\d(\p_n z)},
\end{equation}
where \(\s\) are coordinates on \(\p\Sigma\), the intersection of the brane with the boundary of \(AdS\), and \(p = 2\) or 5 for an M2- or M5-brane, respectively. This implements a Legendre transformation with respect to the position \(\r\) of the end of the brane. There is no need to Legendre transform with respect to the gauge field, as the dissolved M2-brane charge is already fixed by the flux quantization condition \eqref{eq:flux_quantization}.

%%%%%%%%%%%%%%%%%%%%%%%%%%%%%%%%%%%%%%%%%%%%%%%%%%
%%%%%%%%%%%%%%%%%%%%%%%%%%%%%%%%%%%%%%%%%%%%%%%%%%
\subsection{Entanglement entropy and probe branes}
%%%%%%%%%%%%%%%%%%%%%%%%%%%%%%%%%%%%%%%%%%%%%%%%%%
%%%%%%%%%%%%%%%%%%%%%%%%%%%%%%%%%%%%%%%%%%%%%%%%%%

Computing the contribution to entanglement entropy from probe branes using the  Ryu-Takayanagi (RT) prescription~\cite{Ryu:2006bv,Ryu:2006ef} naively requires computing the back-reaction of the brane on the metric, which is usually difficult. However, methods exist which allow the leading order contribution in the probe limit to be obtained without any knowledge of back-reaction~\cite{Jensen:2013lxa, Karch:2014ufa}, based on the techniques of~\cite{Casini:2011kv,Lewkowycz:2013nqa}. We now review the relevant details.

In a conformal field theory in flat space, the entanglement entropy of a spherical entangling region of radius \(R\) is equal to the thermal entropy of that same region after a conformal transformation to hyperbolic space  \cite{Casini:2011kv}. In holography, this transformation is implemented by a diffeomorphism which puts \(AdS\) in hyperbolic slicing, along with an appropriate change in defining function. In terms of the \(AdS_7 \times S^4\) solution \eqref{eq:ads_solution}, the diffeomorphism may be taken to be
\begin{alignat}{2} \label{eq:map_to_hyperbolic}
		t &=  \Omega^{-1} R \sqrt{v^2 - 1} \sinh \tau,
		\qquad
		& & z = \Omega^{-1} R,
		\nonumber \\
		r &= \Omega^{-1} R v \sinh  u \sin \f_0,
		& & x = \Omega^{-1} R v \sinh  u \cos \f_0,
\end{alignat}
where \(\Omega = v \cosh u + \sqrt{v^2 - 1} \cosh \tau\), with other coordinates unchanged. The new coordinates take values in the ranges \(\tau \in (-\infty, \infty)\), \(v \in [1,\infty)\), \(u \in [0,\infty)\), and \(\f_0 \in [0,\pi]\). The gauge field strength \(F_4\) is unchanged under this transformation, while metric becomes
\begin{equation} \label{eq:hyperbolic_slicing}
	\diff s^2 = 4L^2 \le(
		\frac{\diff v^2}{f(v)} - f(v) \diff \tau^2 + v^2 \diff u^2 + v^2 \sinh^2 u \, \diff \f_0^2 + v^2 \sinh^2 u \sin^2 \f_0 \diff s_{S^3}^2
	\ri)
	+ L^2 \diff s_{S^4}^2,
\end{equation}
where \(f(v) = v^2 - 1\).

This coordinate system does not cover all of \(AdS_7\). The pre-image of the coordinate transformation is the entanglement wedge~\cite{Headrick:2014cta}, the causal development of the region at \(t=0\) bounded by the entangling region and the Ryu-Takayanagi surface, given explicitly by \(t^2 + x^2 + r^2 + z^2 \leq R^2\). There is a horizon at \(v=1\), which in flat slicing is the Ryu-Takayanagi surface \(x^2 + r^2 + z^2 = R^2\) at \(t = 0\) \cite{Jensen:2013lxa}. The inverse temperature of the horizon is \(\b_0 = 1/T_0 = 2\pi\).

The metric \eqref{eq:hyperbolic_slicing} and gauge field \eqref{eq:ads_field_strength} remain a solution to the 11D SUGRA equations of motion with the more general metric function
\begin{equation}
	f(v) = v^2 - 1 - \frac{v_H^6 - v_H^4}{v^4}.
\end{equation}
The horizon is now at a position \(v_H\), related to the inverse temperature \(\b\) by
\begin{equation} \label{eq:hyperbolic_temperature}
	v_H = \frac{1}{3\b} \le( \pi + \sqrt{\pi^2 + 6 \b^2} \ri).
\end{equation}

At leading order in the probe limit, the contribution of the brane to the free energy in the dual CFT in hyperbolic space is
\begin{equation} \label{eq:hyperbolic_free_energy}
	F^{(1)}(\b) = \b^{-1} I^\star_\mathrm{brane}(\b),
\end{equation}
where \(I_\mathrm{brane}^\star(\b)\) is the on-shell action of the brane in Euclidean signature,\footnote{We put the metric \eqref{eq:hyperbolic_slicing} in Euclidean signature by a Wick rotation \(\t \to i \tilde \t\). We will abuse notation slightly by dropping the tilde on the Euclidean time coordinate.} with \(\t \sim \t + \b \). The contribution from the brane to the thermal entropy in hyperbolic space, and thus the entanglement entropy in flat space, is therefore given by the thermodynamic identity
\begin{equation} \label{eq:hyperbolic_ee}
	\see^{(1)} = \le. \b^2 \frac{\p F^{(1)}}{\p \b} \ri|_{\b=2\pi}
\end{equation}

One advantage of the probe limit is that in some cases we are able to compute the brane's contribution to the R\'enyi entropies \cite{rényi1961}, which give more information about the reduced density matrix. We define the contribution of the defect to the \(q\)-th R\'enyi entropy, \(S_q^{(1)}\), analogously to its contribution~\eqref{eq:defect_ee} to the entanglement entropy. This may also be computed from the free energy \cite{2011arXiv1102.2098B,Hung:2011nu,Jensen:2013lxa},
\begin{equation} \label{eq:hyperbolic_renyi}
	S_q^{(1)} = \frac{2\pi q}{1 - q} \le[
		F^{(1)}(2\pi) - F^{(1)}(2\pi q)
	\ri].
\end{equation}
In the limit \(q\to1\) this reduces to the entanglement entropy~\eqref{eq:hyperbolic_ee}. Other interesting limits of the R\'enyi entropies are \(q\to0\), which measures the number of non-zero eigenvalues of the reduced density matrix, and \(q\to\infty\), which measures the largest eigenvalue.\footnote{See \cite{Hung:2011nu} for a detailed discussion.}

When the brane embedding breaks conformal symmetry, it is no longer possible to perform the conformal transformation to hyperbolic space. However, the bulk coordinate change to hyperbolic slicing, without changing the defining function, remains a convenient way of computing generalized gravitational entropy~\cite{Lewkowycz:2013nqa}. The contribution of the brane to the entanglement entropy, at leading order in the probe limit, is given by
\begin{equation} \label{eq:generalized_gravitational_entropy}
	\see^{(1)} = 2\pi \lim_{\b \to 2\pi} \p_\b I^\star_{2\pi}(\b),
\end{equation}
where \(I^\star_{2\pi}(\beta)\) is the Euclidean on-shell action of the brane solution in the background~\eqref{eq:hyperbolic_slicing} with arbitrary~\(v_H\), but with the period of Euclidean time given by \(\b_0 = 2\pi\)~\cite{Karch:2014ufa}.

In several of the examples in this paper we were able to find the analytic solution for the brane embedding only at inverse temperature \(\b_0 = 2\pi\), corresponding to pure \(AdS\). We cannot then analytically compute the on-shell action as a function of \(\b\), but we may calculate the entanglement entropy~\eqref{eq:generalized_gravitational_entropy} using the observation of ref.~\cite{Kumar:2017vjv} that the first variation of action with respect to the fields vanishes when the fields satisfy the classical equations of motion. Hence, only \textit{explicit} factors of the inverse temperature contribute to the derivative in \eqref{eq:hyperbolic_ee}, while \textit{implicit} factors of \(\b\) appearing through the dependence of the embedding on the temperature do not. Thus we may choose to take the fields on-shell only after performing the derivative in \eqref{eq:hyperbolic_ee},
\begin{equation} \label{eq:hyperbolic_ee_off_shell}
	\see^{(1)}	= \lim_{\b \to 2\pi} \b \le[ \p_\b I_{2\pi}(\b) \ri]^\star.
\end{equation}
We therefore only need the embedding at inverse temperature \(\b_0 = 2\pi\) to compute the entanglement entropy.

In the hyperbolic slicing, the seven-form field strength is given at all temperatures by
\begin{equation}
	F_7 = 6 (2L)^6 v^5 \sinh^4 u \sin^3 \f_0 \sin^2 \f_1 \sin \f_2 \, \diff \t \wedge \diff v \wedge \diff u \wedge  \diff \f_0 \wedge \diff \f_1 \wedge \diff \f_2 \wedge \diff \f_3. 
\end{equation}
We consider M5-brane embeddings with boundaries. The on-shell action of such an M5-brane may change by boundary terms under gauge transformations of \(C_6\)~\cite{Drukker:2005kx}. The consequence for us is that the entanglement entropy for solutions presented in section~\ref{sec:symmetric} will depend on the choice of gauge for \(C_6\) in the hyperbolic slicing,\footnote{It is plausible that there exists some boundary term which cancels the gauge dependence, but the form of this boundary term is not known to us.} so we must be careful to choose the appropriate gauge. The same phenomenon occurs in the computation of entanglement entropy for defects dual to D3-branes in type IIB SUGRA \cite{Kumar:2017vjv}.

We will choose a gauge which is quite natural given the manifest symmetries of the hyperbolic slicing,\footnote{The authors of \cite{Kumar:2017vjv} chose a gauge which was natural in the Rindler slicing of \(AdS\). We have checked that doing so in our case does not change our results for the entanglement entropy.}
\begin{equation} \label{eq:hyperbolic_c6_gauge}
	C_6 = (2L)^6 (v_H^6 - v^6) \sinh^4 u \sin^3 \f_0 \sin^2 \f_1 \sin \f_2 \, \diff \t \wedge \diff u \wedge  \diff \f_0 \wedge \diff \f_1 \wedge \diff \f_2 \wedge \diff \f_3. 
\end{equation}
Note that this gauge is not the result of performing the coordinate transformation \eqref{eq:map_to_hyperbolic} on the flat slicing gauge potential \eqref{eq:flat_slicing_c6}. In section~\ref{sec:symmetric}, we confirm that with this gauge choice we obtain the same result for the entanglement entropy of a symmetric representation Wilson surface as that computed in ref.~\cite{us_entanglement} using the Ryu-Takayanagi prescription in the fully back-reacted geometry. The latter calculation is independent of the choice of gauge for \(C_6\), so this agreeement supports~\eqref{eq:hyperbolic_c6_gauge} as the correct gauge.

%%%%%%%%%%%%%%%%%%%%%%%%%%%%%%%%%%%%%%%%%%%%%%%%%%
%%%%%%%%%%%%%%%%%%%%%%%%%%%%%%%%%%%%%%%%%%%%%%%%%%
\section{Single M2-brane}
\label{sec:m2}
%%%%%%%%%%%%%%%%%%%%%%%%%%%%%%%%%%%%%%%%%%%%%%%%%%
%%%%%%%%%%%%%%%%%%%%%%%%%%%%%%%%%%%%%%%%%%%%%%%%%%

In this section we compute the entanglement entropy contribution from a single M2-brane, believed to be holographically dual to a Wilson surface operator in the fundamental representation. We begin by reviewing the embedding of the M2-brane in flat slicing \cite{Lunin:2007ab}. 

The M2-brane is described by the action \eqref{eq:m2_action}. We choose static gauge, parameterizing the brane by \(\xi = (t,x,z)\), and take as an ansatz
\(
	r = r(z),
\)
with boundary condition \(\lim_{z\to0}r = 0\). The pullback of \(C_3\) \eqref{eq:c3} onto the brane vanishes with this ansatz, and the action reduces to
\begin{equation} \label{eq:m2_action_ansatz}
	S_\mathrm{M2} = - T_\mathrm{M2} \int \diff t \diff x \diff z \frac{8 L^3}{z^3} \sqrt{1 + r'(z)^2}.
\end{equation}
This is minimised for constant \(r\), so the solution obeying the boundary condition \(r = 0\) at \(z=0\) is
\begin{equation}
	r(z) = 0.
\end{equation}

Substituting this solution into the bulk action, we find that on-shell
\begin{equation}
	S^\star_\mathrm{M2} = -8 T_\mathrm{M2} L^3 \int \diff t \diff x \diff z \frac{1}{z^3} = - \frac{M}{\pi \e^2} \int \diff t \diff x,
\end{equation}
where on the right hand side we have performed the integral over \(z\). Since this integral is UV divergent, we have implemented a cutoff at small \(z = \e\), the same cutoff as used in \eqref{eq:entanglement_general_form}. The boundary term \eqref{eq:boundary_term} turns out to precisely cancel the bulk contribution to the action, so the full on-shell action vanishes,
\begin{equation}
    S^\star_\mathrm{M2} + S^\star_\mathrm{bdy} = 0.
\end{equation}

Mapping to the hyperbolic slicing using \eqref{eq:map_to_hyperbolic}, the solution spans \((\t,v,u)\) and obeys \(\sin\f_0 = 0\). It is straightforward to check that this remains a solution for all temperatures in the hyperbolic slicing. Substituting this solution into the Euclidean action for the M2-brane, we obtain the on-shell action as a function of temperature
\begin{equation} \label{eq:m2_hyperbolic_on_shell_action}
	I_\mathrm{M2}^\star = \frac{4 M}{\pi} \int_0^\b \diff \tau \int_{v_H}^\Lambda \diff v \int_0^{u_c} \diff u \, v
	= \frac{2M}{\pi} \b \le(\Lambda^2 - v_H^2\ri) u_c.
\end{equation}
We have imposed upper limits \(\Lambda\) and \(u_c\) to regulate the integrals over \(v\) and \(u\). The term which diverges as \(\Lambda \to \infty\) is removed by the boundary term \eqref{eq:boundary_term},
\begin{equation} \label{eq:m2_hyperbolic_on_shell_action_boundary}
	I_\mathrm{bdy}^\star = - \frac{2M}{\pi} \b \Lambda^2 u_c.
\end{equation}
The divergence arising from the limit \(u_c \to \infty\) is not removed by this boundary term. In fact, this divergence is physical, and leads to the expected logarithmic divergence in the entanglement entropy. In terms of the cutoff at small \(z=\e\), the large \(u\) cutoff is given by~\cite{Jensen:2013lxa}
\begin{equation} \label{eq:cutoff_identification}
	u_c = \log\le( \frac{2R}{\e} \ri) + \mathcal{O}\le( \frac{\e^2}{R^2} \ri).	 
\end{equation}

Making use of equation \eqref{eq:hyperbolic_temperature} for the position of the horizon, we obtain the contribution of the brane to the free energy as a function of inverse temperature,
\begin{equation} \label{eq:m2_hyperbolic_free_energy}
	F^{(1)}(\b) = \b^{-1}(I_\mathrm{M2}^\star + I_\mathrm{bdy}^\star)= - \frac{2 M}{9 \pi \b^2}\le(
	\pi + \sqrt{\pi^2 + 6 \b^2}
	\ri)^2 u_c.
\end{equation}
Substituting this into \eqref{eq:hyperbolic_renyi}, and identifying the large \(u\) and small \(z\) cutoffs using~\eqref{eq:cutoff_identification}, we find the contribution of the brane to the R\'enyi entropies to be
\begin{equation} \label{eq:m2_renyi}
	S_q^{(1)} = \frac{2}{9} M \frac{1 - 6 q^2 + \sqrt{1 + 24 q^2} }{
		q(1-q)
	}\log\le(
		\frac{2 R}{\e}
	\ri) + \mathcal{O}\le( \frac{\e^2}{R^2} \ri).
\end{equation}
This calculation matches the result in equation (3.33) of \cite{Jensen:2013lxa},\footnote{To obtain our result, set \(d=6\) and \(n=4\) in their formula.} which applies to probe branes in \(AdS\) described by the same bulk action but with different boundary terms. The two calculations agree because the boundary terms are equal when the equations of motion are satisfied.

The entanglement entropy is obtained from the limit \(q \to 1\) of the R\'enyi entropies~\eqref{eq:m2_renyi},
\begin{equation} \label{eq:m2_ee}
	\see^{(1)} = \frac{8}{5} M \log \le(\frac{2 R}{\e} \ri) + \mathcal{O}\le( \frac{\e^2}{R^2} \ri).
\end{equation}
Other physically interesting limits are \(q \to 0\) and \(q \to \infty\),
\begin{subequations}
\begin{align}
	S_{q\to 0}^{(1)} &= \frac{4}{9q} M \log \le( \frac{2 R}{\e} \ri) + \mathcal{O}(q^0),
	\\
	\lim_{q\to\infty} S_{q}^{(1)} &= \frac{4}{3} M \log \le( \frac{2R}{\e} \ri).
\end{align}
\end{subequations}

From the coefficient of the logarithm in the entanglement entropy~\eqref{eq:m2_ee} we obtain the central charge,
\begin{equation} \label{eq:m2_central_charge}
	b = \frac{24}{5} M,
\end{equation}
reproducing the result of \cite{Gentle:2015jma,us_entanglement} in the fundamental representation. This central charge suggests that the number of massless degrees of freedom of a self-dual string scales as \(M\), as opposed to the \(M^3\) scaling of the degrees of freedom in the bulk \(\mathcal{N}=(2,0)\) theory. This is the same scaling found from the chiral R-symmetry anomaly for a single M2-brane stretched between parallel M5-branes~\cite{Berman:2004ew}, as well as coefficients in the defect contribution to the Weyl anomaly~\cite{Graham:1999pm}.

%%%%%%%%%%%%%%%%%%%%%%%%%%%%%%%%%%%%%%%%%%%%%%%%%%
%%%%%%%%%%%%%%%%%%%%%%%%%%%%%%%%%%%%%%%%%%%%%%%%%%
\section{M5-branes wrapping \(S^3 \subset S^4\)}
\label{sec:antisymmetric}
%%%%%%%%%%%%%%%%%%%%%%%%%%%%%%%%%%%%%%%%%%%%%%%%%%
%%%%%%%%%%%%%%%%%%%%%%%%%%%%%%%%%%%%%%%%%%%%%%%%%%

In this section we will seek solutions wrapping an \(S^3\) internal to the \(S^4\) factor of the background geometry. When the world volume of the brane takes the form \(AdS_3 \times S^3\), such a solution is expected to be dual to a Wilson surface in an antisymmetric representation~\cite{Lunin:2007ab}, corresponding to a Young tableau consisting of a single column. The number of boxes \(N\) in the tableau is equal to the amount of M2-brane charge dissolved in the M5-brane, determined from the flux quantization condition \eqref{eq:flux_quantization}.

%%%%%%%%%%%%%%%%%%%%%%%%%%%%%%%%%%%%%%%%%%%%%%%%%%
%%%%%%%%%%%%%%%%%%%%%%%%%%%%%%%%%%%%%%%%%%%%%%%%%%
\subsection{The solution in flat slicing}
%%%%%%%%%%%%%%%%%%%%%%%%%%%%%%%%%%%%%%%%%%%%%%%%%%
%%%%%%%%%%%%%%%%%%%%%%%%%%%%%%%%%%%%%%%%%%%%%%%%%%

Let us parameterize the brane by \(\xi = (t,x,z,\c_1,\c_2,\c_3)\), gauge fix the auxiliary scalar field to \(a = z\), and employ an ansatz
\begin{equation}
    \q = \q(z),
    \quad
    F_3 = \frac{4 L^3 N}{M} \sin^2 \c_1 \sin \c_2 \, \diff \c_1 \wedge \diff \c_2 \wedge \diff \c_3,
\end{equation}
with \(r = 0\). One can verify that this ansatz satisfies the equations of motion for the gauge field. Substituting the ansatz and integrating over \(\c_{1,2,3}\), the PST action~\eqref{eq:pst_action} becomes
\begin{equation} \label{eq:flow_action}
	S_\mathrm{M5} = - \frac{M^2}{4\pi} \int \diff t \diff x \diff z \frac{1}{z^3} \sqrt{
		\le(D(\q)^2 + \sin^6 \q\ri) \le(4 + z^2 \q'^2\ri)
	},
\end{equation}
where
\begin{equation}
	D(\q) = 3 \cos\q - \cos^3 \q - 2 + \frac{4 N}{M}.
\end{equation}

The Euler-Lagrange equation for \(\q\) is
\begin{equation}
	\p_z \le(
		\frac{\q'}{z} \sqrt{
			\frac{ D(\q)^2 + \sin^6 \q}{4 + z^2 \q'^2}
		}
	\ri)
	+ \frac{3 \sin^3 \q}{z^3} \le( D(\q) - \cos \q \sin^2\q \ri) \sqrt{
		\frac{4 + z^2 \q'^2}{D(\q)^2 + \sin^6 \q}
	} = 0.
\end{equation}
This is satisfied by any solution to the first order BPS condition \cite{Camino:2001at,Gomis:1999xs}
\begin{equation} \label{eq:flow_bps}
	\q' = - \frac{2}{z} \frac{
			\p_{\q} \le( D(\q) \cos \q + \sin^4 \q \ri)
		}{
			D(\q) \cos \q + \sin^4 \q
		},
\end{equation}
which ensures that the brane embedding preserves one quarter of the supersymmetries of the background solution.

The BPS condition \eqref{eq:flow_bps} possesses two classes of solutions with constant \(\q\). One class is the antisymmetric Wilson surface~\cite{Lunin:2007ab,Chen:2007ir,Mori:2014tca}, corresponding to a representation of \(\mathfrak{su}(M)\) with a Young tableau consisting of \(N\) boxes. For these solutions, 
\begin{equation} \label{eq:antisymmetric_wilson_surface_angle}
	\cos\q = 1 - 2N/M.
\end{equation} The other class of solution sits at the north or south pole of the \(S^4\), \(\q = 0\) or \(\pi\), with arbitrary \(N\). The wrapped \(S^3\) therefore collapses to zero size and this solution corresponds to a bundle of \(N\) M2-branes~\cite{Gomis:1999xs}.

\begin{figure}
\begin{center}
\begin{subfigure}{0.4\textwidth}
	\begin{center}
	\includegraphics[height=4cm]{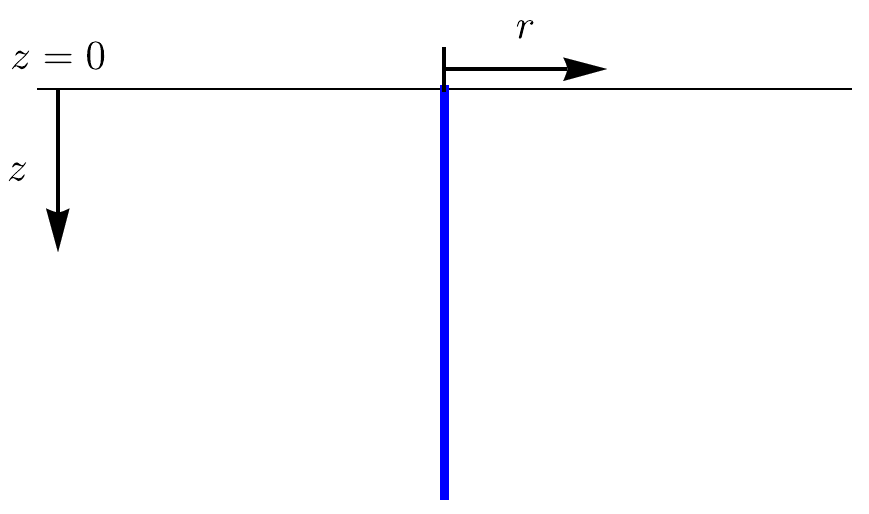}
	\caption{\(AdS_7\)}
	\end{center}
\end{subfigure}
\begin{subfigure}{0.4\textwidth}
	\begin{center}
	\includegraphics[height=4cm]{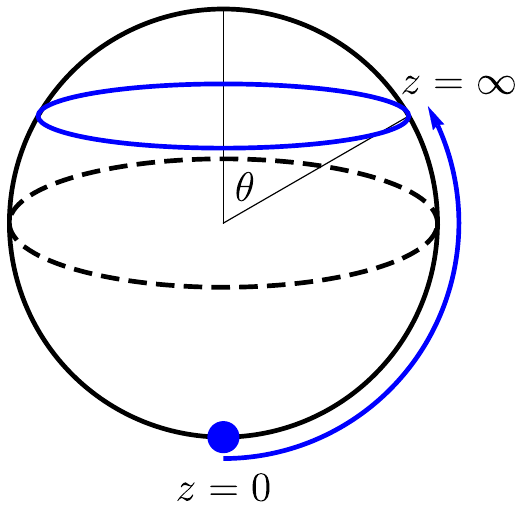}
	\caption{\(S^4\)}
	\end{center}
\end{subfigure}
\caption{Cartoon of the antisymmetric flow embedding in \(AdS_7 \times S^4\). \textbf{(a)}~In the \(AdS_7\) factor of the geometry, the brane (the thick blue line) spans the directions \((t,x,z)\) (the \(t\) and \(x\) directions are suppressed in this figure) and occupies \(r=0\). The thin, horizontal, black line in this figure is the boundary of \(AdS_7\). \textbf{(b)}~In the UV (\(z\to0\)) the M5-brane collapses to the south pole of the \(S^4\).  For non-zero \(z\), the M5-brane wraps an \(S^3\) at a polar angle \(\q(z)\) which decreases with increasing \(z\). In the IR (\(z \to \infty\)) this angle tends to a value~\eqref{eq:antisymmetric_wilson_surface_angle}, determined by the dissolved M2-brane charge. The UV and IR solutions correspond to a bundle of M2-branes and the antisymmetric Wilson surface, respectively.}
\label{fig:antisymmetric_flow_cartoon}
\end{center}
\end{figure}

To obtain solutions where \(\q\) depends non-trivially on \(z\), we integrate the BPS condition~\eqref{eq:flow_bps} to obtain
\begin{equation} \label{eq:antisymmetric_flow_solution}
	\cos \q(z) = c^4 z^4 - \sqrt{
		\le(1 - c^4 z^4 \ri)^2 + \frac{4 N}{M} c^4 z^4
	},
\end{equation}
where \(c\) is an integration constant. This solution tends to each of the constant \(\q\) solutions in opposite limits. In the UV, \(c z \to 0\), \(\q  \to \pi\) and the solution collapses to the bundle of M2-branes. In the IR, \(c z \gg 1\), the solution becomes \(\cos\q \approx 1 - 2N/M\), the antisymmetric Wilson surface. The solution therefore describes an RG flow from the bundle of M2-branes to the antisymmetric Wilson surface, which we will refer to as the antisymmetric flow. We sketch this embedding in figure~\ref{fig:antisymmetric_flow_cartoon}.

Expanding \(\q(z)\) for small \(z\) and using the methods of \cite{deHaro:2000vlm, Karch:2005ms}, we find that the flow is triggered by the vacuum expectation value (VEV) of an operator \(\mathcal{O}_\q\) with conformal dimension \(\Delta = 2\),
\begin{equation}
	\le< \mathcal{O}_\q \ri> = - 2 c^2 \sqrt{\frac{2 N (M - N)}{\pi}}.
\end{equation}

Substituting the BPS condition \eqref{eq:flow_bps} into the action \eqref{eq:flow_action}, the on-shell action may be written in the form
\begin{align} \label{eq:antisymmetric_flow_on_shell_action_bulk}
	S_\mathrm{M5}^\star &= - \frac{M^2}{2 \pi} \int \diff t \diff x \diff z \frac{1}{z^3} \frac{D(\q)^2 + \sin^6 \q}{D(\q) \cos \q + \sin^4\q}
	\nonumber \\
	&= \frac{M^2}{4\pi} \int \diff t \diff x \diff z \, \p_z \le[
		\frac{1}{z^2} \le(
		D(\q) \cos\q + \sin^4 \q
	\ri)
	\ri]
\end{align}
The boundary term \eqref{eq:boundary_term} evaluates to
\begin{equation} \label{eq:antisymmetric_flow_on_shell_action_boundary}
	S_\mathrm{bdy} = \frac{M^2}{4\pi} \int \diff t \diff x \frac{1}{\e^2} \le.\le(
		D(\q) \cos\q + \sin^4 \q
	\ri)\ri|_{z=\e}.
\end{equation}
Noting that the integration over \(z\) in~\eqref{eq:antisymmetric_flow_on_shell_action_bulk} has limits \([\e,\infty)\), and that the contents of the square brackets vanish in the limit \(z \to \infty\), we see that the bulk and boundary contributions cancel. Hence the contribution of the brane to the on-shell action vanishes in flat slicing.

%%%%%%%%%%%%%%%%%%%%%%%%%%%%%%%%%%%%%%%%%%%%%%%%%%
%%%%%%%%%%%%%%%%%%%%%%%%%%%%%%%%%%%%%%%%%%%%%%%%%%
\subsection{Entanglement entropy of the antisymmetric representation Wilson surface}
%%%%%%%%%%%%%%%%%%%%%%%%%%%%%%%%%%%%%%%%%%%%%%%%%%
%%%%%%%%%%%%%%%%%%%%%%%%%%%%%%%%%%%%%%%%%%%%%%%%%%

In this section we compute the entanglement entropy contribution from the M5-brane embedding with constant \(\q = \cos^{-1} \le(1 - 2N/M\ri)\), the antisymmetric Wilson surface.

In hyperbolic slicing, the solution at inverse temperature \(\b_0 = 2\pi\) may be obtained by a coordinate transformation in flat space. It spans \((\t,v,u)\) and satisfies \(\sin \f_0 = 0\). It is straightforward to verify that this is still a solution for arbitrary temperatures in the hyperbolic slicing.

Substituting this solution into the PST action, and Wick rotating to Euclidean signature, we find the bulk contribution to the Euclidean on-shell action to be
\begin{equation} \label{eq:antisymmetric_hyperbolic_action}
	I_\mathrm{M5}^\star = \frac{4 N (M - N)}{\pi} \int_0^\b \diff \tau \int_{v_H}^\Lambda \diff v \int_0^{u_c} \diff u \, v
	= \frac{2N(M - N)}{\pi} \b \le(\Lambda^2 - v_H^2\ri) u_c.
\end{equation}
This is \(N(M-N)/M\) times the result for the M2-brane~\eqref{eq:m2_hyperbolic_on_shell_action}. The same is true for the boundary term,
\begin{equation} \label{eq:antisymmetric_hyperbolic_action_boundary}
	I_\mathrm{bdy}^\star = - \frac{2 N (M-N)}{\pi} \b \Lambda^2 u_c,
\end{equation}
and therefore the contribution from the brane to the free energy in hyperbolic slicing is given by
\begin{equation} \label{eq:antisymmetric_hyperbolic_free_energy}
	F^{(1)}(\b) = - \frac{2 N (M-N)}{9 \pi \b^2} \le(
		\pi + \sqrt{\pi^2 + 6 \b^2}
	\ri)^2 u_c.
\end{equation}

Substituting the free energy into \eqref{eq:hyperbolic_renyi} and using \eqref{eq:cutoff_identification} to relate \(u_c\) an \(\e\), we find that the contribution of the antisymmetric Wilson surface to the \(q\)-th R\'enyi entropy is given by
\begin{equation}
    S_q^{(1)} = \frac{2}{9} N(M-N) \frac{1 - 6 q^2 + \sqrt{1 + 24 q^2} }{q (1-q)} \log \le( \frac{2 R}{\e} \ri).
\end{equation}
Taking the limit \(q \to 1\), the entanglement entropy contribution from the Wilson surface is
\begin{equation} \label{eq:antisymmetric_ee}
    \see^{(1)} = \frac{8}{5} N (M-N) \log \le( \frac{2 R}{\e} \ri).
\end{equation}
In the limits of small and large \(q\), we find respectively
\begin{subequations}
\begin{align}
	S_{q\to 0}^{(1)} &= \frac{4}{9q} N(M - N) \log \le( \frac{2 R}{\e} \ri) + \mathcal{O}(q^0),
	\\
	\lim_{q\to\infty} S_{q}^{(1)} &= \frac{4}{3} N(M - N) \log \le( \frac{2R}{\e} \ri).
\end{align}
\end{subequations}

From the entanglement entropy~\eqref{eq:antisymmetric_ee} we extract the central charge,
\begin{equation} \label{eq:antisymmetric_central_charge}
    b = \frac{24}{5} N (M - N).
\end{equation}
This reproduces the central charge obtained for an antisymmetric representation in \cite{Gentle:2015jma,us_entanglement}. It is invariant under the replacement \(N \to M - N\), corresponding to complex conjugation of the representation of \(\mathfrak{su}(M)\), and reduces to \(N\) times the central charge~\eqref{eq:m2_central_charge} of a single M2-brane for \(N  \ll M\).

%%%%%%%%%%%%%%%%%%%%%%%%%%%%%%%%%%%%%%%%%%%%%%%%%%
%%%%%%%%%%%%%%%%%%%%%%%%%%%%%%%%%%%%%%%%%%%%%%%%%%
\subsection{Entanglement entropy of the antisymmetric flow solution}
\label{sec:antisymmetric_on_shell}
%%%%%%%%%%%%%%%%%%%%%%%%%%%%%%%%%%%%%%%%%%%%%%%%%%
%%%%%%%%%%%%%%%%%%%%%%%%%%%%%%%%%%%%%%%%%%%%%%%%%%

For the flow solution \eqref{eq:antisymmetric_flow_solution}, we have not been able to construct the solution at arbitrary temperature in the hyperbolic slicing. We therefore cannot compute the R\'enyi entropies, but the entanglement entropy may be obtained from equation~\eqref{eq:hyperbolic_ee_off_shell}.

In hyperbolic slicing, we parameterise the brane by \((\t,v,u,\c_1,\c_2,\c_3)\), and gauge fix the auxiliary scalar field to be given by \(a = v\). The embedding will be specified by the function \(\q = \q(\t,v,u)\). As before, the gauge field strength is determined by the flux quantization condition,
\begin{equation}
    F_3 = \frac{4 L^3 N}{M} \sin^2\c_1 \sin\c_2 \, \diff \c_1 \wedge  \diff \c_2 \wedge\diff \c_3.
\end{equation}

Substituting this ansatz into the action and integrating out the wrapped \(S^3\), we find that the Euclidean action for the M5-brane, with arbitrary \(v_H\) but with \(\t \sim \t + 2\pi\), is
\begin{equation} \label{eq:antisymmetric_flow_hyperbolic_action}
    I_{2\pi}(\b) = \int_0^{2\pi} \diff \tau \int_{v_H}^\infty \diff v \int_0^{u_c} \diff u \, \mathcal{L},
\end{equation}
where
\begin{equation} \label{eq:antisymmetric_flow_hyperbolic_lagrangian}
    \mathcal{L} = \frac{M^2}{4\pi} v \sqrt{
        \le( D(\q)^2 + \sin^6 \q \ri)
        \le(
            4 + \frac{1}{f(v)} (\p_\t \q)^2 + f(v) (\p_v \q)^2 + \frac{1}{v^2} (\p_u \q)^2
        \ri)
    }.
\end{equation}
The entanglement entropy is obtained using \eqref{eq:hyperbolic_ee_off_shell}; we differentiate the off-shell action with respect to \(\b\), set \(\b = 2\pi\), and take \(\q\) on-shell.

The resulting integral for the entanglement entropy must be performed numerically. In appendix \ref{app:antisymmetric_flow} we give some details on how we manipulate the integrals into a form suitable for numerical evaluation. As for similar embeddings of D5-branes in \(AdS_5\times S^5\)~\cite{Kumar:2017vjv}, it is convenient to perform a coordinate transformation back to flat slicing, where the embedding is much simpler. The result is that the entanglement entropy is given by\footnote{We use \(x^0\) to denote the Euclidean time coordinate in flat slicing.}
\begin{align} \label{eq:antisymmetric_flow_ee}
    \see^{(1)} &= \frac{2 M^2}{5} \int_\e^R \diff z \frac{R}{z \sqrt{R^2 - z^2}} \frac{D(\q)^2 + \sin^6\q}{D(\q) \cos\q + \sin^4\q}
    \nonumber \\ &\phantom{=}
    - \frac{8 M^2 R^4}{5\pi} \int_{z\geq\e} \diff x^0 \diff x \diff z \frac{z\mathbf{N}_1}{\mathbf{D}_1} \frac{
        \le( D(\q) \sin \q - \cos \q \sin^3 \q \ri)^2
        }{
        D(\q) \cos\q + \sin^4 \q
    },
\end{align}
where \(\q\) is given by the solution \eqref{eq:antisymmetric_flow_solution}, and
\begin{subequations} \label{eq:antisymmetric_flow_ee_factors}
    \begin{align}
       \mathbf{N}_1 &=
            \left[R^4+2 R^2 \left(x_0-x\right) \left(x+x_0\right)+\left(x^2+x_0^2\right)^2\right]^2
            \nonumber \\ &\phantom{=}
            -2 z^4 \left[R^4+R^2 \left(6
            x_0^2-2 x^2\right)+x^4+6 x^2 x_0^2+5 x_0^4\right]
            \nonumber \\ &\phantom{=}
            -4 x_0^2 z^2 \left[(R-x)^2+x_0^2\right] \left[(R+x)^2+x_0^2\right]-4
            x_0^2 z^6+z^8,
       \\
       \mathbf{D}_1 &=
            \left[(R-x)^2+x_0^2+z^2\right]^2 \left[(R+x)^2+x_0^2+z^2\right]^2
            \nonumber \\ &\phantom{=}
            \times \left[R^4+2 R^2 \left(-x^2+x_0^2-z^2\right)+\left(x^2+x_0^2+z^2\right)^2\right]^2.
    \end{align}
\end{subequations}

The entanglement entropy \eqref{eq:antisymmetric_flow_ee} is logarithmically divergent at small \(\e\),
\begin{equation}
    \see^{(1)} = \frac{8}{5} M (M - N) \log \le( \frac{2R}{\e} \ri) + \mathcal{O}(\e^0).
\end{equation}
The divergent term is the entanglement entropy of the UV solution, namely \((M-N)\) times the entanglement entropy~\eqref{eq:m2_ee} of a single M2-brane. We will obtain a UV finite quantity by subtracting this contribution, yielding the difference \(\Delta \see^{(1)}\) between the entanglement entropy of the flow solution and the bundle of M2-branes,
\begin{equation}
    \Delta \see^{(1)} = \see^{(1)} - \frac{8}{5} M (M - N) \log \le( \frac{2 R}{\e} \ri).
\end{equation}

\begin{figure}[t!]
	\begin{subfigure}{0.5\textwidth}
		\includegraphics[width=\textwidth]{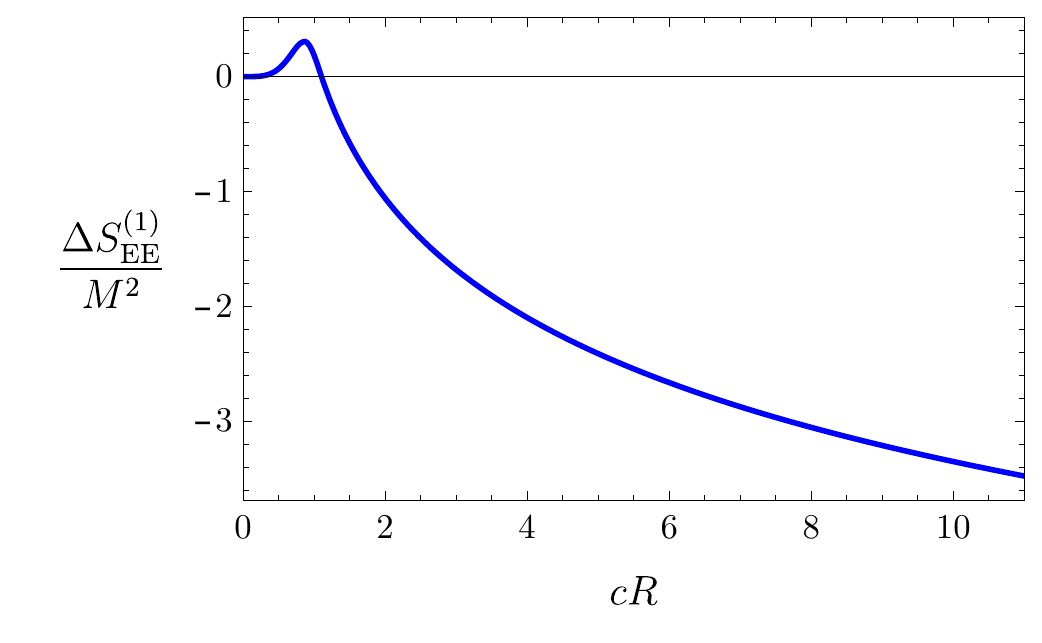}
		\caption{\(\dfrac{N}{M} = \dfrac{1}{10}\)}
	\end{subfigure}
	\begin{subfigure}{0.5\textwidth}
		\includegraphics[width=\textwidth]{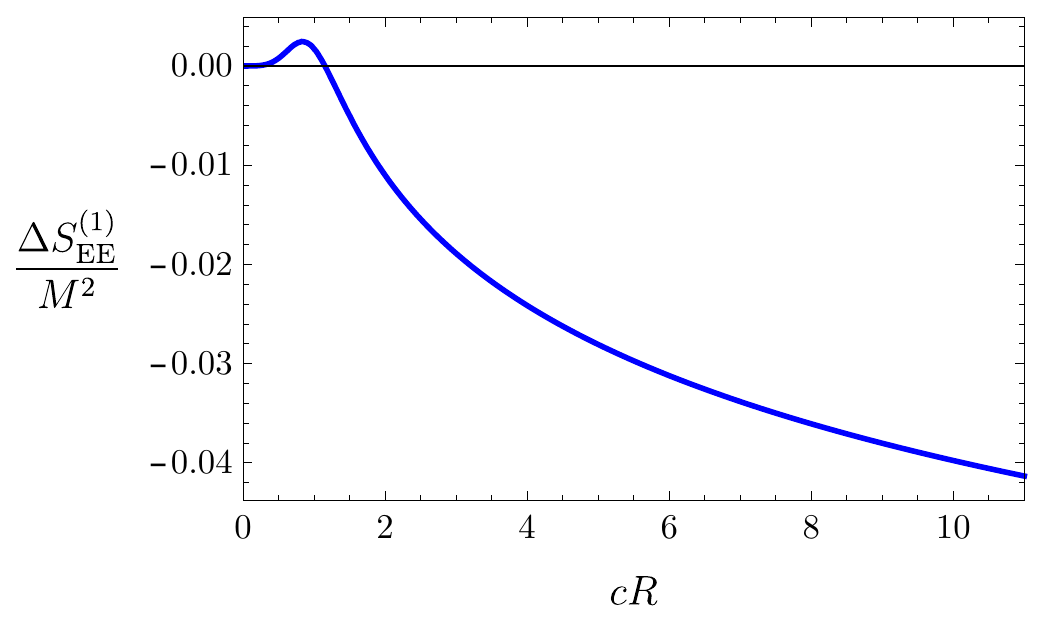}
		\caption{\(\dfrac{N}{M} = \dfrac{9}{10}\)}
	\end{subfigure}
	\\[1em]
	\begin{subfigure}{0.5\textwidth}
		\includegraphics[width=\textwidth]{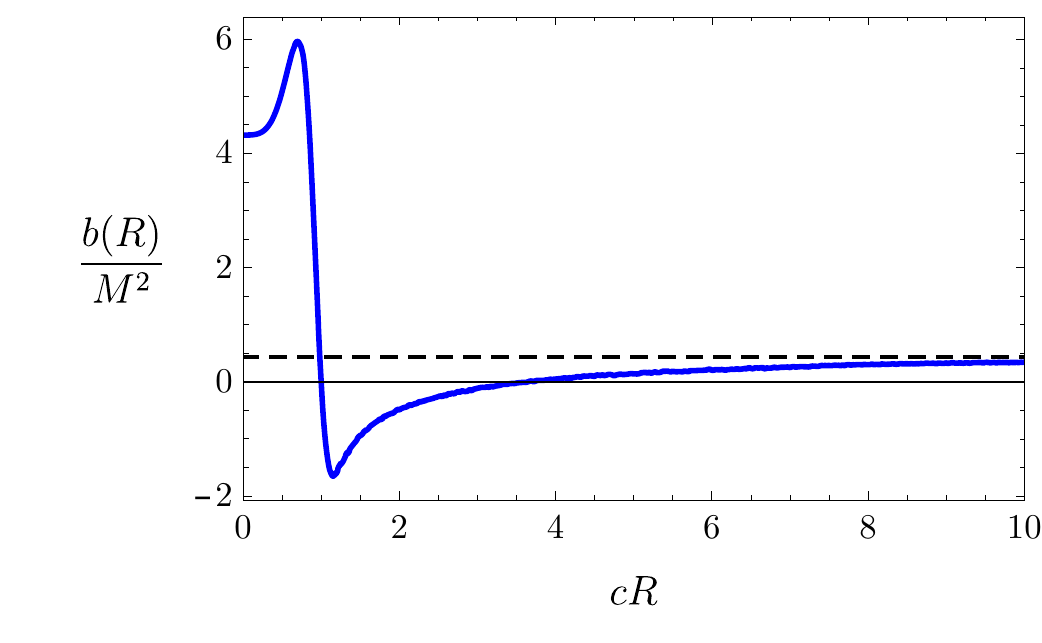}
		\caption{\(\dfrac{N}{M} = \dfrac{1}{10}\)}
	\end{subfigure}
	\begin{subfigure}{0.5\textwidth}
		\includegraphics[width=\textwidth]{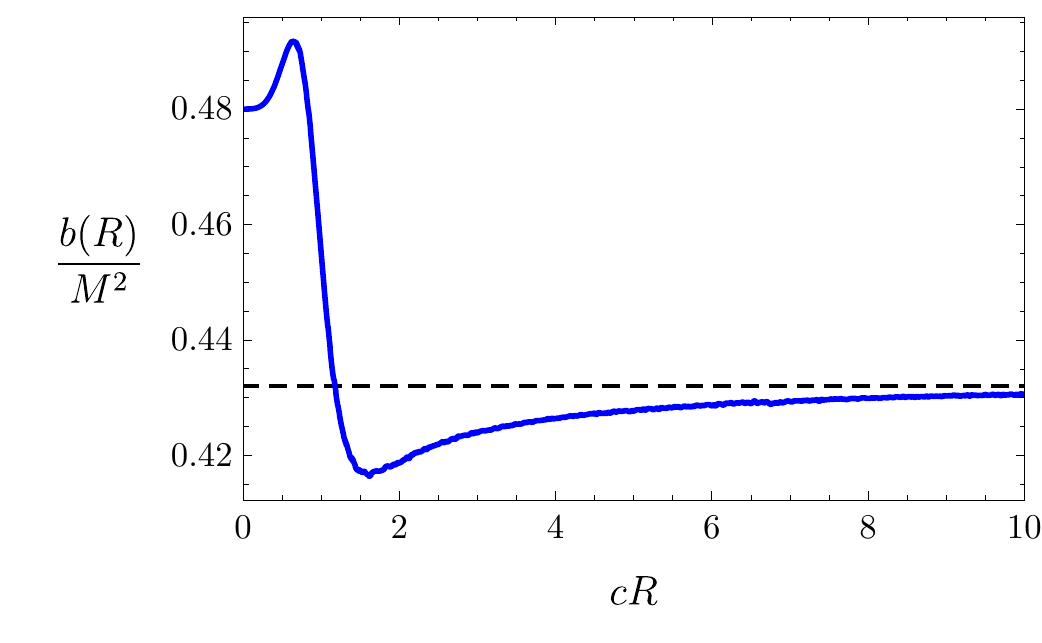}
		\caption{\(\dfrac{N}{M} = \dfrac{9}{10}\)}
	\end{subfigure}
	\caption{Numerical results for the entanglement entropy (top row) and \(b\)-function (bottom row) defined in~\eqref{eq:b_function}, for the defect RG flow from a bundle of \(N\) M2-branes in the UV to the antisymmetric Wilson surface in the IR. For small values of \(cR\), the \(b\)-function is given by the \((M-N)\) times the central charge for a single M2-brane~\eqref{eq:m2_central_charge}. As \(cR \to \infty\), the \(b\)-function tends to the central charge of the antisymmetric Wilson surface~\eqref{eq:antisymmetric_central_charge}, indicated by the horizontal dashed lines in the plots.}
	\label{fig:antisymmetric_flow_ee}
\end{figure}
In figure~\ref{fig:antisymmetric_flow_ee} we plot our numerical results for \(\Delta \see^{(1)}\) and the \(b\)-function \eqref{eq:b_function}, both as functions of the dimensionless combination \(cR\). The difference in  entanglement entropy, \(\Delta \see^{(1)}\), vanishes in the limit \(cR \to 0\), by definition. Increasing \(cR\) from zero, \(\Delta \see^{(1)}\) at first increases, before reaching a maximum and then decreasing, apparently without bound.

For \(cR \to 0\), the \(b\)-function tends to the central charge of the UV solution, namely \(N\) times the central charge for a single M2-brane \eqref{eq:m2_central_charge}, \(b(R=0) \equiv b_\mathrm{UV} = \frac{24}{5} M (M - N)\). Similarly, for \(cR \to \infty\), the \(b\)-function tends to the central charge of the IR solution --- the antisymmetric Wilson surface. Thus \(\lim_{R\to\infty} b(R) \equiv b_\mathrm{IR} = \frac{24}{5} N (M - N)\). The \(b\)-function interpolates between these two limits non-monotonically.

Since these flows only exist for \(N \leq M\), the central charge is manifestly larger in the UV than in the IR,
\begin{equation} \label{eq:b_inequality}
	b_\mathrm{UV} \geq b_\mathrm{IR}.
\end{equation}
This would appear to support the interpretation of \(b\) as a measure of the massless degrees of freedom on the brane. However, in section~\ref{sec:symmetric} we will see that for M5-brane flow solutions wrapping an \(S^3\) internal to \(AdS_7\) instead of \(S^4\), the inequality \eqref{eq:b_inequality} does not hold.

%%%%%%%%%%%%%%%%%%%%%%%%%%%%%%%%%%%%%%%%%%%%%%%%%%
%%%%%%%%%%%%%%%%%%%%%%%%%%%%%%%%%%%%%%%%%%%%%%%%%%
\subsection{On-shell action}
%%%%%%%%%%%%%%%%%%%%%%%,%%%%%%%%%%%%%%%%%%%%%%%%%%%
%%%%%%%%%%%%%%%%%%%%%%%%%%%%%%%%%%%%%%%%%%%%%%%%%%

It has recently been argued \cite{Kobayashi:2018lil} that for defect RG flows the free energy on a sphere or in hyperbolic space serves as a better candidate \(C\)-function than the entanglement entropy. For the bundle of M2-branes in hyperbolic slicing, the free energy is \((M-N)\) times that of a single M2-brane, given in~\eqref{eq:m2_hyperbolic_free_energy}, while the free energy of the antisymmetric Wilson surface was computed in \eqref{eq:antisymmetric_hyperbolic_free_energy}. Setting \(\b = \b_0 = 2\pi\), we find
\begin{equation}
	- F^{(1)}(\b_0) = \begin{cases}
		\dfrac{2}{\pi} M(M-N) u_c, \quad &\text{for the bundle of M2-branes},
		\\[1em]
		\dfrac{2}{\pi} N(M-N) u_c, \quad &\text{for the antisymmetric Wilson surface}.
	\end{cases}
\end{equation}
Since these flows only exist for \(N \leq M\), we find that \(-F^{(1)}\) is indeed larger in the UV than the IR, consistent with the expectations of \cite{Kobayashi:2018lil}.

However, since the antisymmetric flow is triggered by a VEV rather than a source, the flow and the bundle of M2-branes describe two different states of the same theory. As pointed out in ref.~\cite{Kobayashi:2018lil}, the difference between the hyperbolic space free energies of the IR and UV solutions is equal to the relative entropy of the two states, and is guaranteed to be positive due to monotonicity of relative entropy \cite{Blanco:2013joa}. 

A candidate \(C\)-function which we can study along the entire flow is provided by ref.~\cite{Kumar:2017vjv}, in which the contribution of probe D-brane solutions to the on-shell action in the entanglement wedge was observed to vary monotonically along a defect RG flow. We will now test whether the same is true for the antisymmetric M5-brane flow.

The entanglement wedge on-shell action, which we will denote by \(S_\mathcal{W}^\star\), is given by~\eqref{eq:antisymmetric_flow_on_shell_action_bulk} and \eqref{eq:antisymmetric_flow_on_shell_action_boundary} but with the domain of integration replaced by the region
\begin{equation}
	\mathcal{W}'_\e = \{t^2 + x^2 +z^2 \leq R^2\} \cap \{z \geq \e\}.
\end{equation}
This is the intersection of the probe brane with the entanglement wedge, with the cutoff region at \(z < \e\) excised. Explicitly
\begin{equation} \label{eq:antisymmetric_flow_entanglement_wedge_action}
	S_\mathcal{W}^\star = \frac{M^2}{2\pi} \int_{\mathcal{W}'_\e} \diff x^0 \diff x \diff z \frac{1}{z^3} \frac{D(\q)^2 + \sin^6 \q}{D(\q) \cos\q + \sin^4 \q}
	- \frac{M^2}{4\pi} \int_{\p \mathcal{W}_\e'} \diff t \diff x \frac{1}{\e^2} \le[D(\q) \cos\q + \sin^4\q\ri]_{z=\e},
\end{equation}
where \(\p\mathcal{W}_\e'\) is the region of the boundary of \(\mathcal{W}_\e'\) intersecting the cutoff surface at \(z = \e\). Note that we have assumed that there are no boundary terms arising from the change in the bulk 11D SUGRA action due to the back-reaction of the brane.

The action \eqref{eq:antisymmetric_flow_entanglement_wedge_action} diverges logarithmically as \(\e\to 0\). For the solutions which preserve defect conformal symmetry, we find
\begin{equation} \label{eq:antisymmetric_flow_action_fixed_points}
	S_\mathcal{W}^\star = \begin{cases}
		2M(M-N) \log \le(\dfrac{R}{\e}\ri) + \mathcal{O}(\e^1), \quad &\text{for the bundle of M2-branes},
		\\[1em]
		2N(M-N)\log \le(\dfrac{R}{\e}\ri) + \mathcal{O}(\e^1), \quad &\text{for the antisymmetric Wilson surface}.
	\end{cases}
\end{equation}
For the flow solution, the coefficient of the logarithmic divergence is the same as for the bundle of M2-branes, but the \(\mathcal{O}(\e^0)\) term will be different.

\begin{figure}[t!]
	\begin{subfigure}{0.5\textwidth}
		\includegraphics[width=\textwidth]{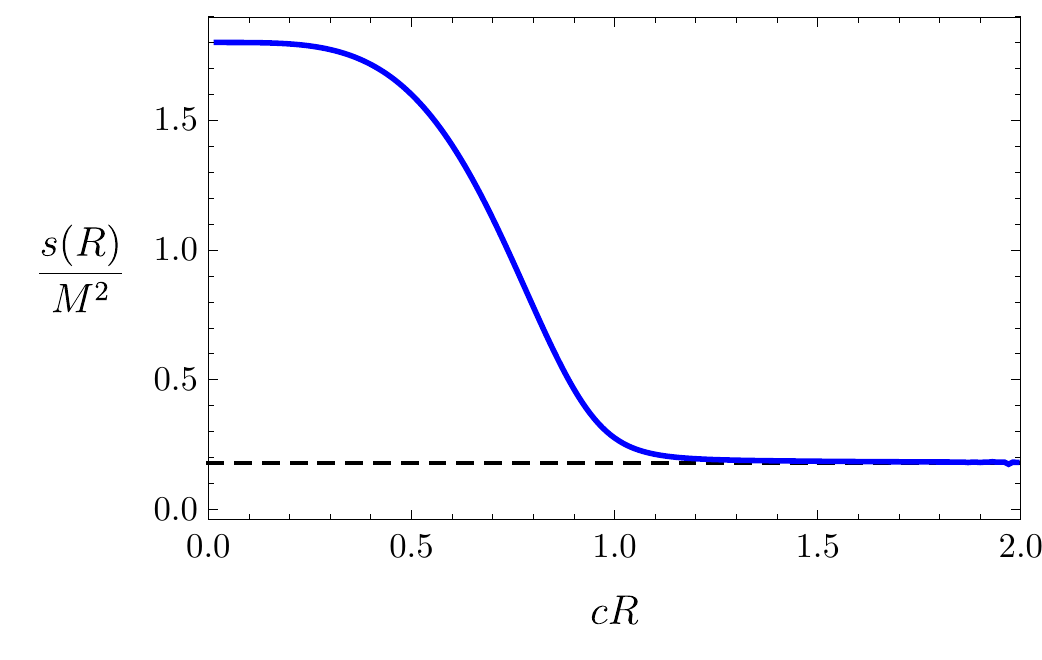}
		\caption{\(\dfrac{N}{M} = \dfrac{1}{10}\)}
	\end{subfigure}
	\begin{subfigure}{0.5\textwidth}
		\includegraphics[width=\textwidth]{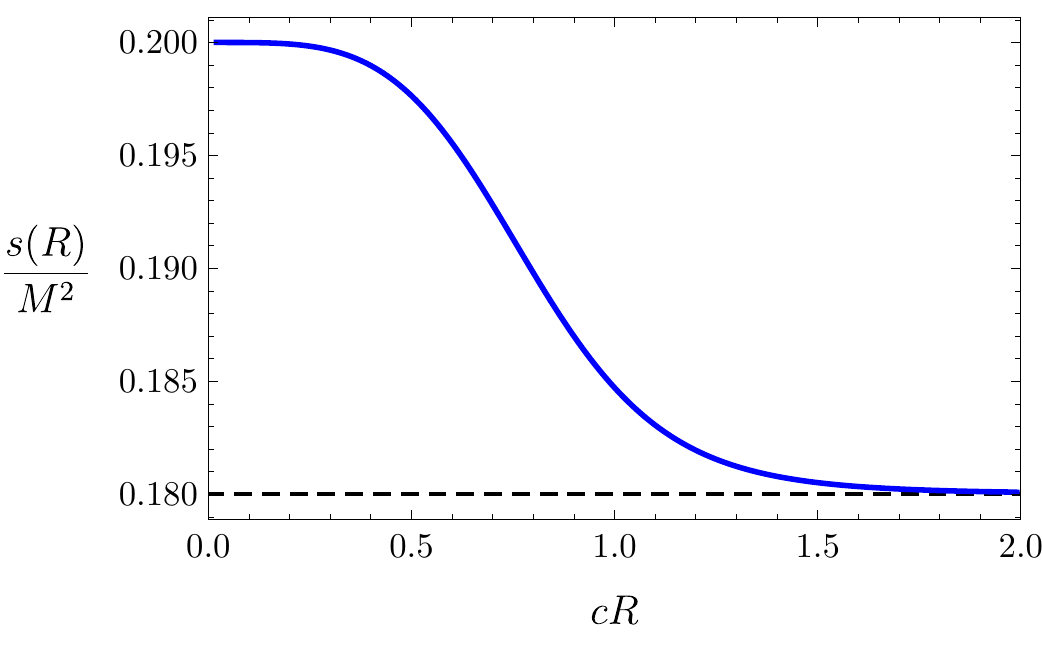}
		\caption{\(\dfrac{N}{M} = \dfrac{9}{10}\)}
	\end{subfigure}
	\caption{Numerical results for the function \(s(R)\), defined in equation~\eqref{eq:s_definition} as minus the logarithmic derivative with respect to \(R\) of the brane's contribution to the on-shell action in the entanglement wedge of a spherical subregion of radius \(R\). For small \(cR\) the derivative tends to the coefficient of the logarithmic divergence of the on-shell action for the UV solution, the bundle of M2-branes. For large \(cR\) the derivative tends to the coefficient for the IR solution, the antisymmetric Wilson surface (indicated by the horizontal dashed lines). In between these limits, \(s(R)\) decreases monotonically.}
	\label{fig:antisymmetric_flow_action}
\end{figure}
For the flow, we must evaluate~\eqref{eq:antisymmetric_flow_entanglement_wedge_action} numerically. To obtain a UV finite quantity we take a logarithmic derivative with respect to the sphere radius \(R\), defining a function
\begin{equation} \label{eq:s_definition}
	s(R) \equiv R\frac{\diff S_\mathcal{W}^\star}{\diff R}.
\end{equation}
Figure~\ref{fig:antisymmetric_flow_action} shows our results, for two sample values of \(N/M\). 

In the limits \(R \to 0\) or \(\infty\), \(s(R)\) tends to the values at the UV or IR fixed points, respectively, given by the coefficients of the logarithms in~\eqref{eq:antisymmetric_flow_action_fixed_points}. Since the flows only exist for \(N \leq M\), this implies that \(s\) is smaller in the IR than in the UV. For all values of \(N/M\) that we have checked, \(s(R)\) appears to decrease monotonically along the flow.

%%%%%%%%%%%%%%%%%%%%%%%%%%%%%%%%%%%%%%%%%%%%%%%%%%
%%%%%%%%%%%%%%%%%%%%%%%%%%%%%%%%%%%%%%%%%%%%%%%%%%
\section{M5-branes wrapping \(S^3 \subset AdS_7\)}
\label{sec:symmetric}
%%%%%%%%%%%%%%%%%%%%%%%%%%%%%%%%%%%%%%%%%%%%%%%%%%
%%%%%%%%%%%%%%%%%%%%%%%%%%%%%%%%%%%%%%%%%%%%%%%%%%

In this section we will seek solutions wrapping an \(S^3\) internal to the \(AdS_7\) factor of the background geometry. This includes the symmetric representation Wilson surface~\cite{Lunin:2007ab}, corresponding to a Young tableau consisting of a single row of \(N\) boxes, with \(N\) determined from~\eqref{eq:flux_quantization}. We will also study flows from the symmetric representation Wilson surface to a bundle of M2-branes.

%%%%%%%%%%%%%%%%%%%%%%%%%%%%%%%%%%%%%%%%%%%%%%%%%%
%%%%%%%%%%%%%%%%%%%%%%%%%%%%%%%%%%%%%%%%%%%%%%%%%%
\subsection{The solution in flat slicing}
%%%%%%%%%%%%%%%%%%%%%%%%%%%%%%%%%%%%%%%%%%%%%%%%%%
%%%%%%%%%%%%%%%%%%%%%%%%%%%%%%%%%%%%%%%%%%%%%%%%%%

We begin by working in a supergravity background of the form \eqref{eq:metric_with_h}, leaving the function \(h(\r)\) arbitrary. We employ static gauge on the brane, \(\xi = (x^0, x^1, x^\a)\), where \(\a\) runs from 2 to 5. With the following ansatz for the world volume fields
\begin{equation}
	\r = \r(x^\a),
	\quad
	F_{3,\a\b\g} = \e_{\a\b\g\d} \d^{\d\e} \h_\e(x^\a),
	\quad
	a = a(x^1),
\end{equation}
we find that the Euler-Lagrange equations for the world volume fields are satisfied if \(\h_\a = \p_\a \r\) and
\begin{equation}
	\d^{\a\b} \p_\a \p_\b \r(x^\a) = 0.
\end{equation}
This is just the four-dimensional flat-space Laplace equation, so we find an infinite family of solutions
\begin{equation} \label{eq:multi_center_spike}
	\r = \r_0 + \frac{2 L^3}{M} \sum^{n}_{a=1} \frac{N_a}{\d^{\a\b} \le(x_\a - y^{(a)}_\a\ri)\le(x_\b - y^{(a)}_\b\ri)}
\end{equation}
With constants \(\r_0\), \(N_a\) and \(y^{(a)}_\a\) determined by the boundary conditions.

Such solutions are well known in flat space, they describe an M5-brane at \(\r=\r_0\), with \(n\) infinite tension self-dual strings with \(N_a\) units of charge at positions \(y^{(a)}\) \cite{Howe:1997ue}. The solution \eqref{eq:multi_center_spike} is the generalization for a probe M5-brane embedded in the geometry~\eqref{eq:metric_with_h} produced by a stack of parallel M5-branes, as studied in~\cite{Gauntlett:1999xz}.

Let us now take the near horizon limit, so that \(h(\r) = L^3/\r^3\), and consider the case \(n = 1\), \(N_1 = N\), and \(y^{(1)} = 0\). The solution \eqref{eq:multi_center_spike} reduces to
\begin{equation}
	\r = \r_0 + \frac{2 L^3 N}{M r^2},
\end{equation}
where \(r = \sqrt{\d_{\a\b} x^\a x^\b}\). Substituting \(\r = L^3/z^2\) and solving for \(r\), we obtain the embedding in the \(AdS_7 \times S^4\) metric~\eqref{eq:ads_solution},
\begin{equation} \label{eq:spike_solution}
	r(z) = \sqrt{ \frac{N}{2M} }\frac{z}{\sqrt{1 + \ct z^2}},
\end{equation}
where \(\ct = - \r_0/4L^3\). The corresponding field strength is given by
\begin{equation} \label{eq:spike_field_strength}
	F_3 = \frac{4L^3 N}{M} \sin^2 \f_1 \sin \f_2 \, \diff \f_1 \wedge \diff \f_2 \wedge \diff \f_3.
\end{equation}

Substituting the solution~\eqref{eq:spike_solution} into the full bulk action~\eqref{eq:spike_action_ansatz} for the brane, we find that the on-shell PST action in flat slicing is
\begin{equation}
	S_\mathrm{M5}^\star = - \frac{M N}{\pi \e^2} \int \diff t \diff x.
\end{equation}
As for the antisymmetric flow solution, this is completely cancelled by the boundary term~\eqref{eq:boundary_term}, so the renormalized on-shell action in flat slicing vanishes.

\begin{figure}[t!]
\begin{center}
\begin{subfigure}{0.32\textwidth}
	\includegraphics[width=\textwidth]{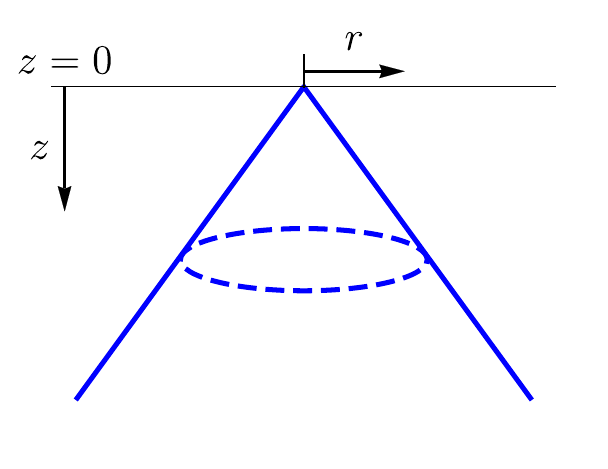}
	\caption{\(\tilde{c} = 0\): Wilson surface}
\end{subfigure}
\begin{subfigure}{0.32\textwidth}
	\includegraphics[width=\textwidth]{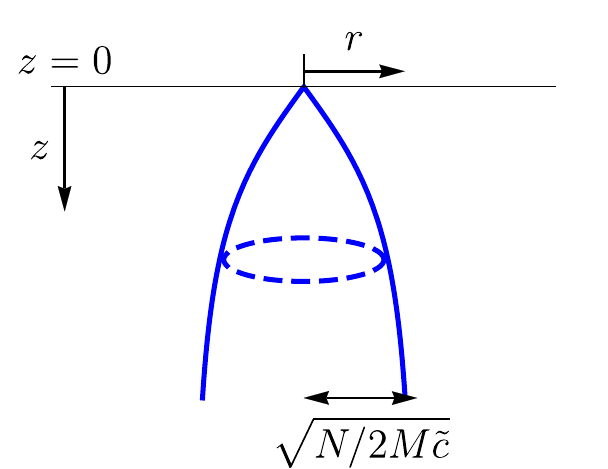}
	\caption{\(\tilde{c} > 0\): Symmetric flow}
\end{subfigure}
\begin{subfigure}{0.32\textwidth}
	\includegraphics[width=\textwidth]{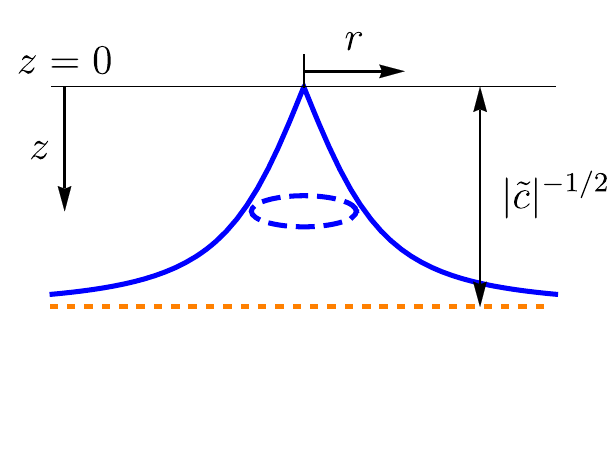}
	\caption{\(\tilde{c} < 0\): Funnel}
\end{subfigure}
\caption{Cartoons of the different M5-brane embeddings wrapping an \(S^3\) internal to \(AdS_7\). In each case the horizontal black line denotes the boundary of \(AdS_7\) at \(z = 0\), and the dashed blue line denotes the \(S^3\) wrapped by the brane. \textbf{(a)} For \(\tilde{c} = 0\) the brane is dual to a symmetric representation Wilson surface. \textbf{(b)} For \(\tilde{c} > 0\) the brane describes a defect RG flow from a symmetric representation Wilson surface to a bundle of M2-branes. In terms of the coordinate \(r\), as \(z \to \infty\) the radius of the wrapped \(S^3\) tends to a finite value \(r = \sqrt{N/2 M \ct}\), so the proper radius vanishes in this limit. \textbf{(c)} For \(\tilde{c} = 0\) the solution is a funnel, created by M2-branes ending on a Coulomb branch M5-brane at \(z = |\tilde{c}|^{-1/2}\).}
\label{fig:symmetric_cartoon}
\end{center}
\end{figure}

The interpretation of the solution depends on the sign of \(\ct\), as sketched in figure~\ref{fig:symmetric_cartoon}. When \(\ct = 0\), the induced metric on the M5-brane world volume is
\begin{equation}
	\diff s_\mathrm{M5}^2 = \frac{4L^2}{z^2} \le[
		-\diff t^2 + \diff x^2 + \le(1 + \frac{N}{2M}\ri) \diff z^2
	\ri]
	+ \frac{2 L^2 N}{M} \diff s_{S^3}^2,
\end{equation}
which is the metric of \(AdS_3 \times S^3\), where the radius of the \(AdS_3\) is \(2L \sqrt{1 + N/2M}\) and the radius of the \(S^3\) is \(L\sqrt{2 N/M}\). The presence of the \(AdS_3\) indicates that the defect preserves the global subgroup of two-dimensional conformal symmetry. Indeed, the solution with \(\ct=0\) is expected to be holographically dual to a Wilson surface in a symmetric representation \cite{Lunin:2007ab,Chen:2007ir,Mori:2014tca}.\footnote{See also ref.~\cite{Chen:2007tt} for a similar M5-brane embedding in \(AdS_4 \times S^7\)}.

When \(\ct \neq 0\) the M5-brane world volume no longer includes an \(AdS_3\) factor, so the defect conformal symmetry is broken. Near the boundary, where \(|\ct| z^2 \ll 1\), the solution approaches the Wilson surface solution. For \(\ct < 0\), \(r(z)\) becomes infinite at a finite value \(z=|\ct|^{-1/2}\). We interpret this solution as a Coulomb branch brane at \(z=|\ct|^{-1/2}\), probed by an infinite tension self-dual string. We will refer to this as the M5-brane funnel.

For \(\ct > 0\), \(r(z)\) remains finite for all \(z\). In the infrared, \(\ct z^2 \gg 1\) the world volume again has an \(AdS_3\) factor but with radius \(2L\), indicating that the solution with positive \(\ct\) is dual to a defect RG flow. At large \(z\) the induced metric takes the form
\begin{equation}
	\diff s_\mathrm{M5}^2 = \frac{4 L^2}{z^2} \le(-\diff t^2 + \diff x^2 + \diff z^2 \ri) + \frac{2 L^2 N}{M \ct z^2} \diff s_{S^3}^2 + \ldots\;,
\end{equation}
where the dots indicate corrections of higher order in \(\ct z^2\). The proper radius of \(S^3\) shrinks to zero as \(z\to\infty\), and a natural guess is that the infrared is a bundle of M2-branes. Similar D3-brane solutions in \(AdS_5 \times S^5\), flowing from a symmetric representation Wilson surface to a bundle of strings, were studied in \cite{Kumar:2016jxy,Kumar:2017vjv}. Expanding the solution \eqref{eq:spike_solution} for small \(z\), we find that as in the case of the flow involving the antisymmetric representation the flow is triggered by the VEV of an operator \(\le<\mathcal{O}_r\ri>\) with conformal dimension \(\Delta = 2\). Explicitly
\begin{equation}
	\le< \mathcal{O}_r \ri> = - \ct N \sqrt{ \frac{M}{\pi ( M + N /2)} }.
\end{equation}

To support the intuition that the infrared is a bundle of non-interacting M2-branes, let us carry out a calculation in the style of section 2.4 of \cite{Kobayashi:2006sb}. Substituting the field strength~\eqref{eq:spike_field_strength} into the M5-brane action, along with the ansatz \(r = r(z)\), we obtain
\begin{equation} \label{eq:spike_action_ansatz}
	S_\mathrm{M5} = - \frac{4 M^2}{\pi} \int \diff t \diff x \diff z \frac{1}{z^6} \le[
		\sqrt{
			\le(\frac{N^2}{4 M^2}z^6  + r^6\ri)
			\le(1 + r'^2\ri)
			} - r^3 r'
	\ri].
\end{equation}
As \(z \to \infty\), \(r\) remains finite and \(r' \to 0\) when evaluated on the solution \eqref{eq:spike_solution}. To leading order at large \(z\), we may therefore neglect the \(r^3 r'\) term compared to the square root, and the \(r^6\) term inside the square root. Thus for large \(z\)
\begin{equation}
	S_\mathrm{M5} \approx - \frac{2 M N}{\pi} \int \diff t \diff x \diff z \frac{1}{z^3} \sqrt{1 + r'^2}.
\end{equation}
This is \(N\) times the action~\eqref{eq:m2_action_ansatz} for a single M2-brane, as expected. Note that in the calculation of ref.~\cite{Kobayashi:2006sb} it was necessary to Legendre transform the action with respect to the gauge field to fix the total amount of fundamental string charge dissolved in the brane. In our case there is no need to Legendre transform, since the M2-brane charge is already fixed by the flux quantization condition~\eqref{eq:flux_quantization}.

%%%%%%%%%%%%%%%%%%%%%%%%%%%%%%%%%%%%%%%%%%%%%%%%%%
%%%%%%%%%%%%%%%%%%%%%%%%%%%%%%%%%%%%%%%%%%%%%%%%%%
\subsection{Entanglement entropy of the symmetric representation Wilson surface}
%%%%%%%%%%%%%%%%%%%%%%%%%%%%%%%%%%%%%%%%%%%%%%%%%%
%%%%%%%%%%%%%%%%%%%%%%%%%%%%%%%%%%%%%%%%%%%%%%%%%%

In the hyperbolic slicing of \(AdS_7\), the solution \eqref{eq:spike_solution} with \(\ct = 0\) becomes
\begin{equation} \label{eq:symmetric_wilson_surface_hyperbolic}
	\Phi \equiv \sin \f_0 = \frac{\kappa}{v \sinh u},
\end{equation}
where \(\kappa^2 \equiv N/2M\). We have not been able to analytically find the generalization of this solution for arbitrary temperatures of the horizon in the hyperbolic slicing. This means we cannot compute the contribution of the Wilson surface to the R\'enyi entropies, but we may still obtain the entanglement entropies by differentiating the off-shell action with respect to the inverse temperature and using \eqref{eq:hyperbolic_ee_off_shell}.

To write the off-shell action, we parameterize the brane by \(\xi = (\t,v,u,\f_1,\f_2,\f_3)\) and take as an ansatz \(\Phi = \Phi(\t,v,u)\). Substituting this into the PST action and integrating over the \(S^3\) parameterized by \((\f_1,\f_2,\f_3)\), we obtain
\begin{equation}
	I_\mathrm{M5} = \int_0^{2\pi} \diff \t \int_{v_H}^\infty \diff v \int_{u_\mathrm{min}}^{u_c} \diff u \, \mathcal{L},
\end{equation}
where
\begin{align} \label{eq:symmetric_off_shell_lagrangian}
	\mathcal{L} &= \frac{8M^2}{\pi} (1 - \Phi^2)^{-1/2} \biggl[v  \le(
		\kappa^4 + \Phi^6 v^6 \sinh^6 u
	\ri)^{1/2} \biggl(
		1 - \Phi^2 + v^2 \sinh^2 u \, \hat{g}^{ab} \p_a \Phi \p_b \Phi
	\biggr)^{1/2}
	\nonumber \\ &\phantom{= \frac{8M^2}{\pi} (1 - \Phi^2)^{-1/2} \biggl[ }
	+ (v^6 - v_H^6) \sinh^4 u \Phi^3 \p_v \Phi\biggr].
\end{align}
The metric \(\hat{g}\) is defined such that
\begin{equation}
	\hat{g}^{ab} \p_a \Phi \p_b \Phi = \frac{1}{f(v)} (\p_\t \Phi)^2 + f(v) (\p_v \Phi)^2
	+ \frac{1}{v^2} (\p_u \Phi)^2.
\end{equation}
When \(v_H=1\) this is the metric of unit-radius \(AdS_3\).

The lower limit \(u_\mathrm{min}\) on the integration over \(u\) is a function of \(v\), determined by the requirement that \(\sin^2\f_0 \leq 1\). From the solution~\eqref{eq:symmetric_wilson_surface_hyperbolic}, we find
\begin{equation}
	\sinh u_\mathrm{min}(v) = \frac{\kappa}{v}.
\end{equation}
Differentiating the off-shell action with respect to \(\b\), taking the limit \(\b \to 2\pi\), and substituting the solution \eqref{eq:symmetric_wilson_surface_hyperbolic}, we find that the contribution to the entanglement entropy from the symmetric representation Wilson surface is given by the integral
\begin{align}
	\see^{(1)} &= \frac{4}{5} N (N + 2M) \int_{u_\mathrm{min}(1)}^{u_c} \diff u \frac{\sinh u}{\sqrt{\sinh^2 u - \kappa^2}}
	\nonumber \\ & \phantom{=}
	+ 4 N^2 \int_1^\infty \diff v \int_{u_\mathrm{min}(v)}^{u_c} \diff u \frac{\sinh u}{\sqrt{v^2 \sinh^2 u - \kappa^2}}.
\end{align}

Performing the integrals, and identifying the cutoff \(u_c\) with the small \(z\) cutoff \(\e\) using~\eqref{eq:cutoff_identification}, we find
\begin{equation}
	\see^{(1)} = \frac{8}{5} N \le( M - \frac{N}{8} \ri) \log \le(
		\frac{2R}{\e}
	\ri) + \mathcal{O}(\e^0),
\end{equation}
reproducing the result of~\cite{Gentle:2015jma,us_entanglement} for the symmetric representation. From this we obtain the central charge for a Wilson surface with representation determined by a Young tableau consisting of a single row of \(N\) boxes:
\begin{equation} \label{eq:symmetric_central_charge}
	b = \frac{24}{5} N \le(M - \frac{N}{8}\ri).
\end{equation}
This matches the appropriate limit of the results of~\cite{Gentle:2015jma,us_entanglement}. This is true even when in the limit \(N \gg M\), in which the probe limit is unreliable. The only requirement is that the Young tableau is a single row. The central charge vanishes at a critical value \(N = 8M\), and is negative for larger \(N\). We have not observed anything else special about this particular value of \(N\).

%%%%%%%%%%%%%%%%%%%%%%%%%%%%%%%%%%%%%%%%%%%%%%%%%%
%%%%%%%%%%%%%%%%%%%%%%%%%%%%%%%%%%%%%%%%%%%%%%%%%%
\subsection{Entanglement entropy of the non-conformal solutions}
%%%%%%%%%%%%%%%%%%%%%%%%%%%%%%%%%%%%%%%%%%%%%%%%%%
%%%%%%%%%%%%%%%%%%%%%%%%%%%%%%%%%%%%%%%%%%%%%%%%%%

We now compute the entanglement entropy contribution from the solutions with \(\ct \neq 0\). We leave the details of the computation to appendix \ref{app:spike_entanglement}. The final result is that the entanglement entropy is given by the integral
\begin{align} \label{eq:spike_entanglement_integral}
	\see^{(1)} &= 
	\frac{8 M N}{5} \int_{\e}^{z_*} \diff z \frac{
		\sqrt{1 + \ct z^2}
		}{
			z \sqrt{1 - (1 + \kappa^2 - \ct R^2) z^2 - \ct z^4 / R^2}
		} \le(
			1 + \frac{\kappa^2}{(1 + \ct z^2)^3}
		\ri)
	\nonumber \\ &\phantom{=}
	- \frac{8 M^2}{5 \pi} \int_{z\geq\e} \diff x^0 \diff x \diff z \frac{8 R^6 r^6 \mathbf{N}_2 }{
		\kappa^2 z^3 \le[
			\le(R^2 - x_0^2 - x^2 - r^2 - z^2 \ri)^2 + 4 R^2 \le(x_0^2 + z^2\ri)
		\ri]^3
	},
\end{align}
where \(r\) is the solution \eqref{eq:spike_solution} and \(z_*\) is the value of the radial coordinate \(z\) at the intersection between the brane and the RT surface, given explicitly by
\begin{equation} \label{eq:zmax}
	z_*^2 = \frac{1}{2|\ct|} \le[
		\sqrt{\le(1 + \frac{N}{2M} - \ct R^2\ri)^2 + 4 \ct R^2} - \le(1 + \frac{N}{2M} - \ct R^2\ri)
	\ri].
\end{equation}
The factor \(\mathbf{N}_2\) appearing in the last integral in~\eqref{eq:spike_entanglement_integral} is given by
\begin{align} \label{eq:spike_entanglement_coeff}
	\mathbf{N}_2 &= 
	\frac{
		4 x_0^2 z^2 (r-\kappa  z)^2 (r+\kappa  z)^2 \left[\left(r^2-R^2+x^2+x_0^2+z^2\right)^2+4 R^2 \left(x_0^2+z^2\right)\right]
	}{
		\left[2 x_0^2 \left(r^2+R^2+x^2+z^2\right)+\left(r^2-R^2+x^2+z^2\right)^2+x_0^4\right]^2
	}
	\nonumber \\ &\phantom{=}
	+\frac{
		6 r^2 \left[2 \kappa ^2 z^4 \left(r^2-R^2+x^2+x_0^2+z^2\right)+r^2 \left(r^2-R^2+x^2+x_0^2\right)^2+4 r^2 R^2 x_0^2-r^2 z^4\right]
	}{
		\left(r^2-R^2+x^2+x_0^2+z^2\right)^2+4 R^2 x_0^2
	}
	\nonumber \\ &\phantom{=}
	-\frac{\left[2 \kappa ^2 z^4 \left(r^2-R^2+x^2+x_0^2+z^2\right)+r^2 \left(r^2-R^2+x^2+x_0^2\right)^2+4 r^2 R^2 x_0^2-r^2 z^4\right]^2}{\left[\left(r^2-R^2+x^2+x_0^2+z^2\right)^2+4 R^2 x_0^2\right]^2}.
\end{align}

\begin{figure}
\begin{subfigure}{0.5\textwidth}
    \includegraphics[width=\textwidth]{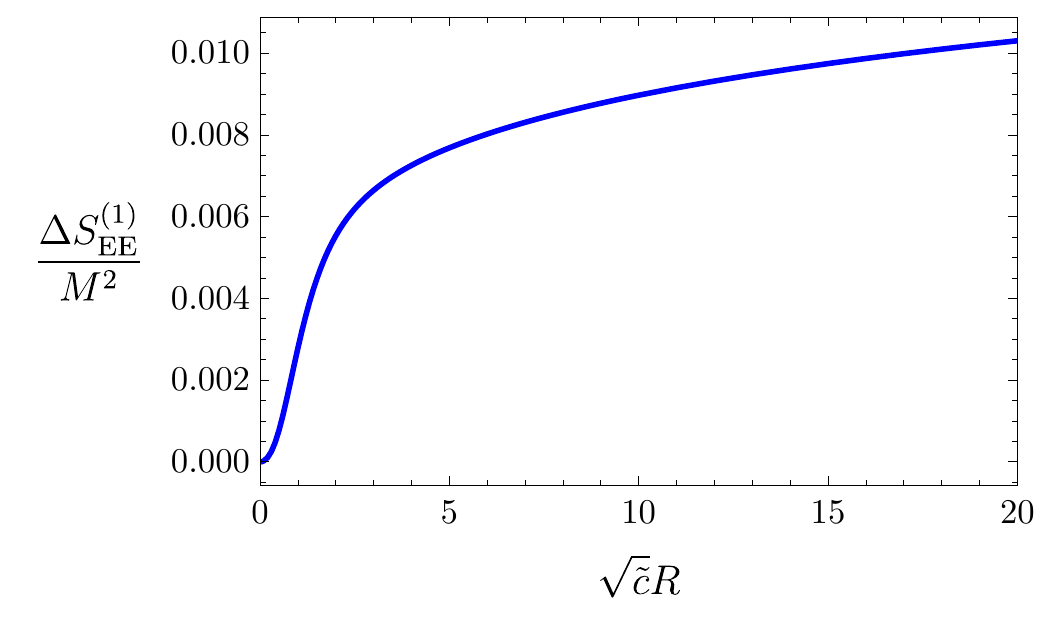}
    \caption{\(\ct > 0,~\dfrac{N}{M} = \dfrac{1}{10}\)}
\end{subfigure}
\begin{subfigure}{0.5\textwidth}
    \includegraphics[width=\textwidth]{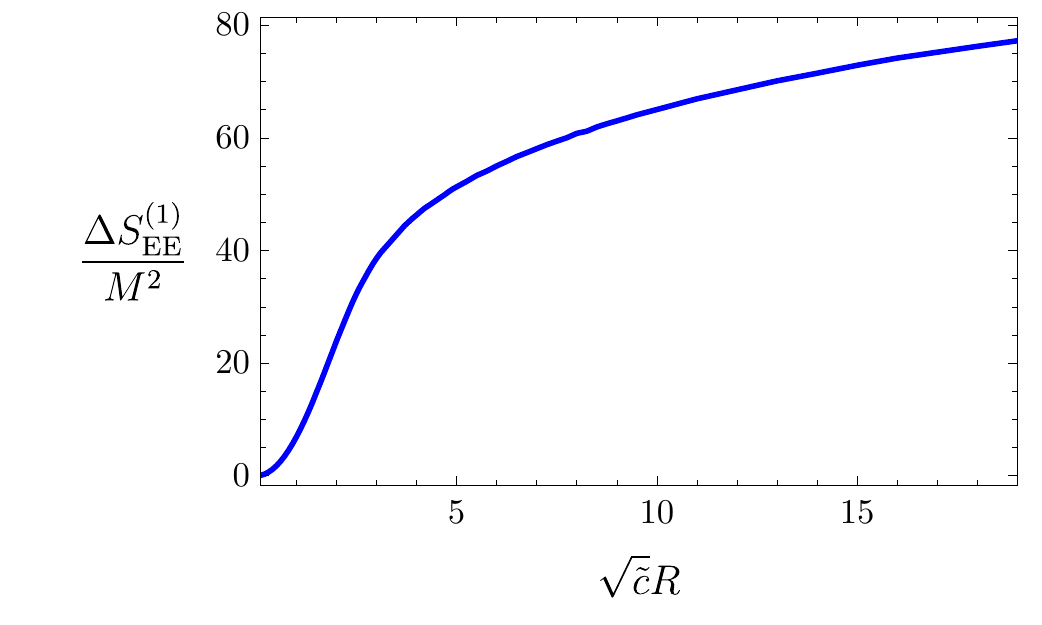}
    \caption{\(\ct > 0,~\dfrac{N}{M} = 10\)}
\end{subfigure}
\\[1em]
\begin{subfigure}{0.5\textwidth}
    \includegraphics[width=\textwidth]{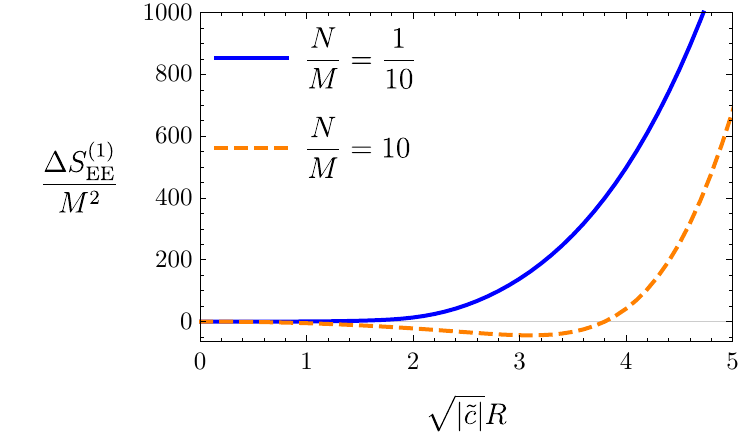}
    \caption{\(\ct < 0\)}
\end{subfigure}
\begin{subfigure}{0.5\textwidth}
    \includegraphics[width=\textwidth]{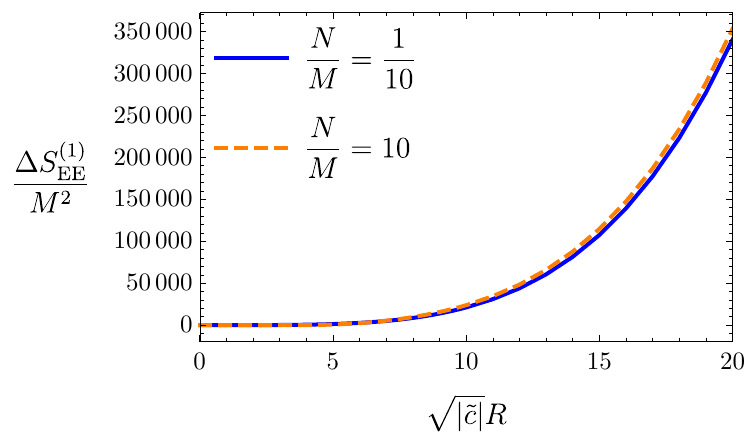}
    \caption{\(\ct < 0\), larger \(R\)}
\end{subfigure}
\caption{
    	The defect contribution to entanglement entropy as a function of the radius \(R\) of the entangling region, for the M5-brane solutions which wrap an \(S^3\) internal to \(AdS_7\). To obtain a UV finite quantity we have calculated the difference between the entanglement entropy of the full solution and that of the symmetric representation Wilson surface. For either sign of \(\ct\) the entanglement entropy appears to grow without bound at large \(\sqrt{|\ct|} R\).
	}    
	\label{fig:spike_entanglement}
\end{figure}

Taking the limit \(\sqrt{|\ct|}R \to 0\), the formula \eqref{eq:spike_entanglement_integral} for the entanglement entropy reduces to the entanglement entropy of a symmetric representation Wilson surface. For non-vanishing \(\ct\), evaluating the integral \eqref{eq:spike_entanglement_integral} requires numerics. As for the antisymmetric flow solutions, we obtain a finite quantity by subtracting the UV contribution, obtaining the excess due to the flow
\begin{equation}
	\Delta \see^{(1)} = \see^{(1)} - \le.\see^{(1)}\ri|_\mathrm{symmetric}.
\end{equation}
Numerical results for \(\Delta \see^{(1)}\), for both signs of \(\ct\) and sample values \(N/M\), are plotted in figure~\ref{fig:spike_entanglement}.

\begin{figure}
    \begin{subfigure}{0.5\textwidth}
        \includegraphics[width=\textwidth]{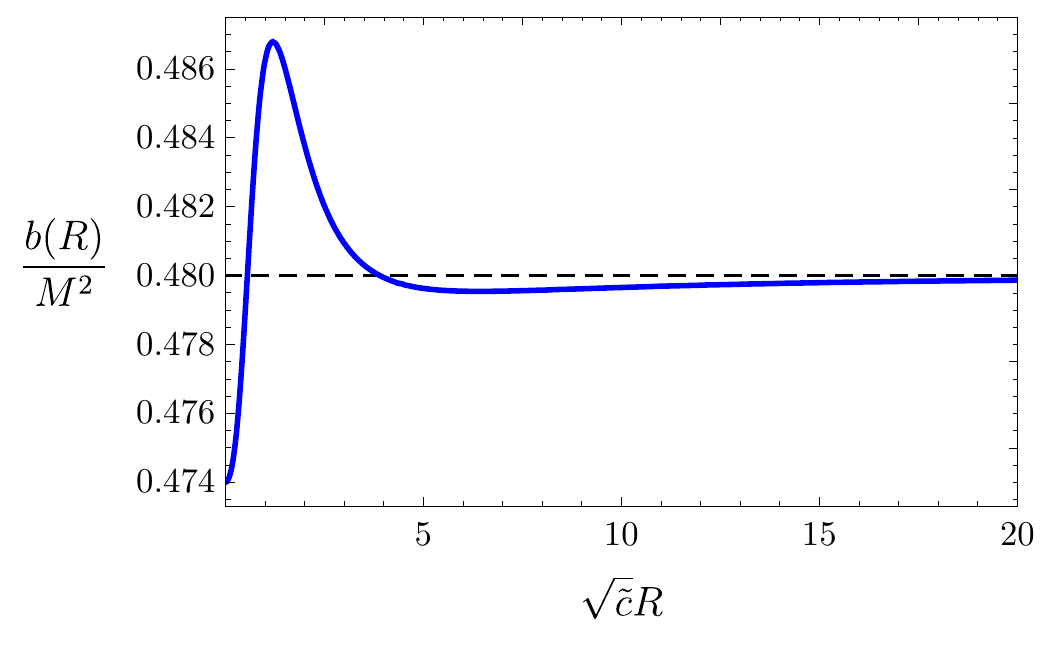}
        \caption{\(\ct > 0,~\dfrac{N}{M} = \dfrac{1}{10}\)}
    \end{subfigure}
    \begin{subfigure}{0.5\textwidth}
        \includegraphics[width=\textwidth]{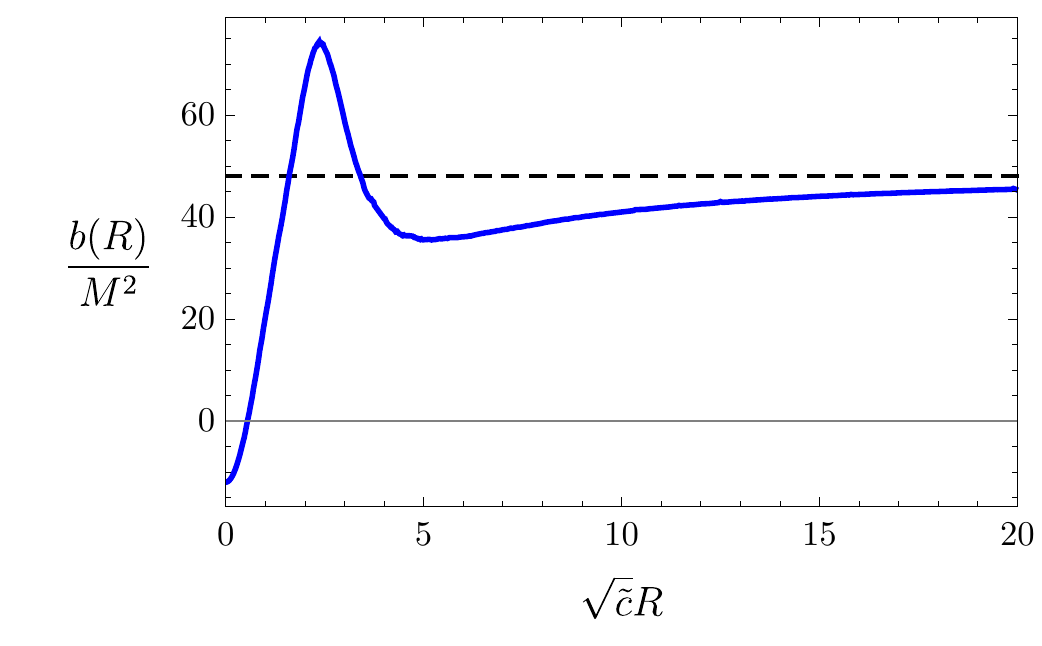}
        \caption{\(\ct > 0,~\dfrac{N}{M} = 10\)}
	\end{subfigure}
	\\[1em]
    \begin{subfigure}{0.5\textwidth}
        \includegraphics[width=\textwidth]{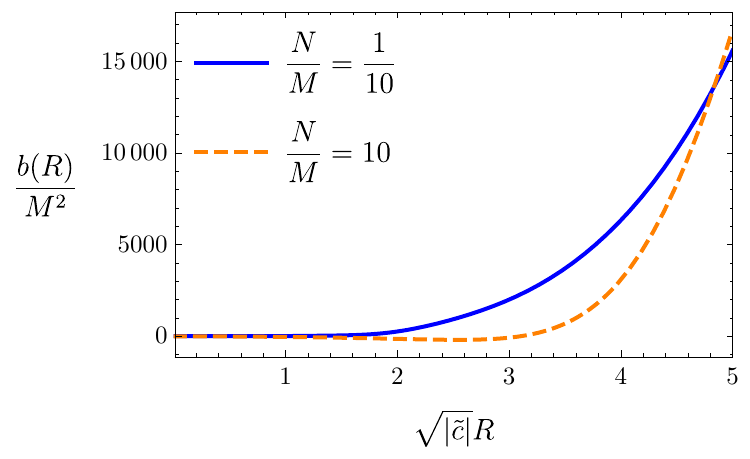}
        \caption{\(\ct < 0\)}
    \end{subfigure}
    \begin{subfigure}{0.5\textwidth}
        \includegraphics[width=\textwidth]{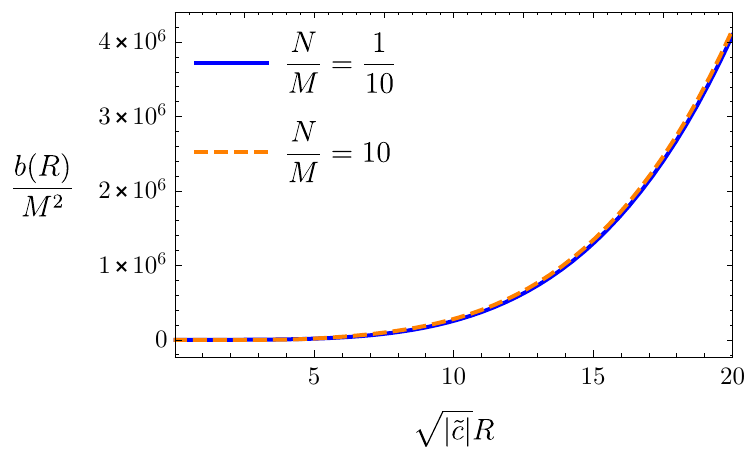}
        \caption{\(\ct < 0\), larger \(R\)}
    \end{subfigure}
    \caption{
		The \(b\)-function for solutions wrapping \(S^3 \subset AdS_7\). For both signs of \(\ct\), \(b(R)\) tends to the UV central charge \eqref{eq:symmetric_central_charge} as \(\sqrt{|\ct|} R \to 0\). For \(\ct>0\), dual to a defect RG flow, as \(\sqrt{\ct}R \to \infty\) the \(b\)-function tends toward the IR value \eqref{eq:symmetric_flow_ir_central_charge}, as indicated by the horizontal dashed line. The \(b\)-function interpolates non-monotonically between these two limits. For \(\ct < 0 \) the \(b\)-function appears to increase without bound.
        \label{fig:spike_b_function}
    }    
\end{figure}
From the entanglement entropy, we compute the \(b\)-function as defined in \eqref{eq:b_function}. The numerical results are shown in figure \ref{fig:spike_b_function}. In the limit \(\sqrt{|\ct|}R \to 0\), \(b(R)\) is given by the central charge of the symmetric representation Wilson surface~\eqref{eq:symmetric_central_charge}. When \(\ct>0\), the solution flows to a bundle of \(N\) M2-branes in the IR. In the limit \(\sqrt{\ct} R \to 1\) the \(b\)-function approaches the expected infrared value of the central charge, namely \(N\) times the central charge of a fundamental representation Wilson surface,
\begin{equation} \label{eq:symmetric_flow_ir_central_charge}
	b_\mathrm{IR} = \frac{24}{5} M N.
\end{equation}
This is greater than the value in the UV \eqref{eq:symmetric_central_charge}. When \(\ct < 0\), \(b\)-function appears to increase without bound for large \(\ct R\). 

%%%%%%%%%%%%%%%%%%%%%%%%%%%%%%%%%%%%%%%%%%%%%%%%%%
%%%%%%%%%%%%%%%%%%%%%%%%%%%%%%%%%%%%%%%%%%%%%%%%%%
\subsection{On-shell action}
%%%%%%%%%%%%%%%%%%%%%%%%%%%%%%%%%%%%%%%%%%%%%%%%%%
%%%%%%%%%%%%%%%%%%%%%%%%%%%%%%%%%%%%%%%%%%%%%%%%%%

We now repeat the analysis of section~\ref{sec:antisymmetric_on_shell} for the solutions wrapping an \(S^3\) internal to \(AdS_7\). In the UV, these solutions tend to the symmetric representation Wilson surface for which the on-shell action in hyperbolic slicing is determined by substituting the solution~\eqref{eq:symmetric_wilson_surface_hyperbolic} into the action \eqref{eq:symmetric_off_shell_lagrangian}. Explicitly, we find
\begin{equation}
	- F^{(1)} = \frac{2}{\pi} N \le( M + \frac{N}{4} \ri) u_c + \mathcal{O}(u_c^0).
\end{equation}
The infrared of the symmetric flow is a bundle of \(N\) M2-branes, with free energy given by \(N\) times the free energy~\eqref{eq:m2_hyperbolic_free_energy} of a single M2-brane. Thus \(- F^{(1)}\) is larger in the UV than in the IR, as guaranteed by monotonicity of relative entropy.

We now turn to the evaluation of the contribution of the probe brane to the Lorentzian signature on-shell action inside the entanglement wedge. This is given by the integral
\begin{equation}
	S_\mathcal{W}^\star = - \frac{2 M N}{\pi} \int_{\mathcal{W}''_\e} \diff t \diff x \diff z \frac{1}{z^3}
	+ \frac{M N}{\pi \e^2} \int_{\p \mathcal{W}''_\e} \diff t \diff x.
\end{equation}
The domain of integration is restricted to the cutoff entanglement wedge
\begin{equation}
	\mathcal{W}''_\e = \{t^2 + x^2 + r^2(z) +z^2 \leq R^2\} \cap \{z \geq \e\},
\end{equation}
with \(r(z)\) given by the solution \eqref{eq:spike_solution}, and \(\p\mathcal{W}''_\e\) denotes the part of the boundary of this surface at \(z=\e\).

The integrals may be evaluated explicitly to give
\begin{equation}
	S_\mathcal{W}^\star = 2 N \le( M + \frac{N}{2} \ri)
		\log \le(
		\frac{z_*}{\e}
	\ri)
	+ \frac{1}{2} N \left[-N \log \left(1 + \ct z_*^2\right)+2 M \left(\frac{R^2}{z_*^2}-1\right)-N\right] + \mathcal{O}(\e),
\end{equation}
where \(z_*\) is the maximal value of \(z\) inside the entanglement wedge, given by~\eqref{eq:zmax}.

\begin{figure}[t!]
	\begin{subfigure}{0.5\textwidth}
		\includegraphics[width=\textwidth]{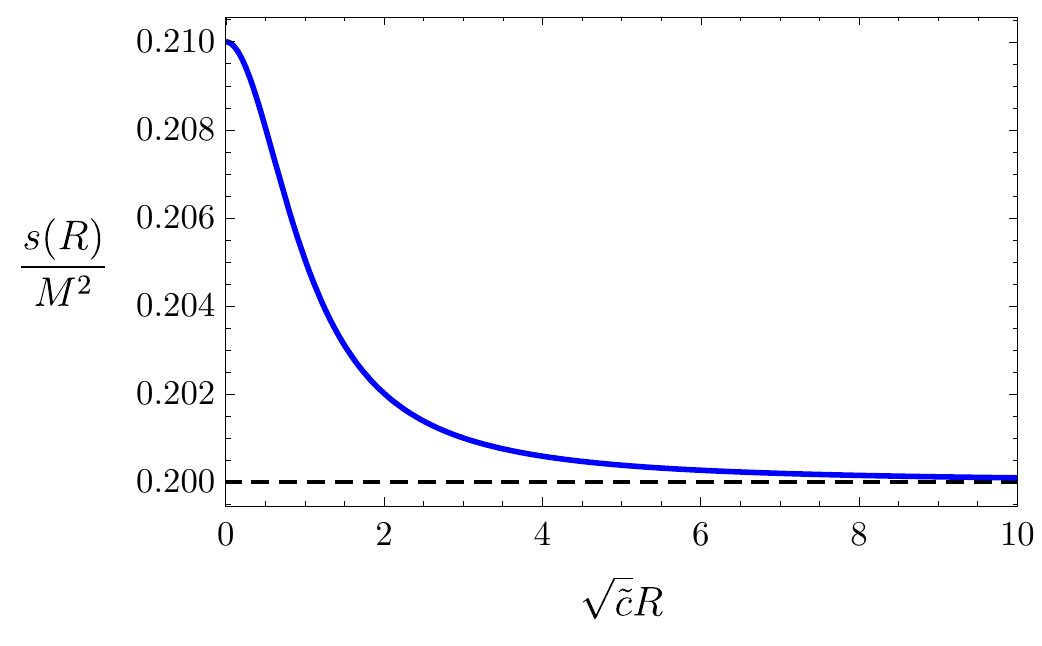}
		\caption{\(\ct > 0,~\dfrac{N}{M} = \dfrac{1}{10}\)}
	\end{subfigure}
	\begin{subfigure}{0.5\textwidth}
		\includegraphics[width=\textwidth]{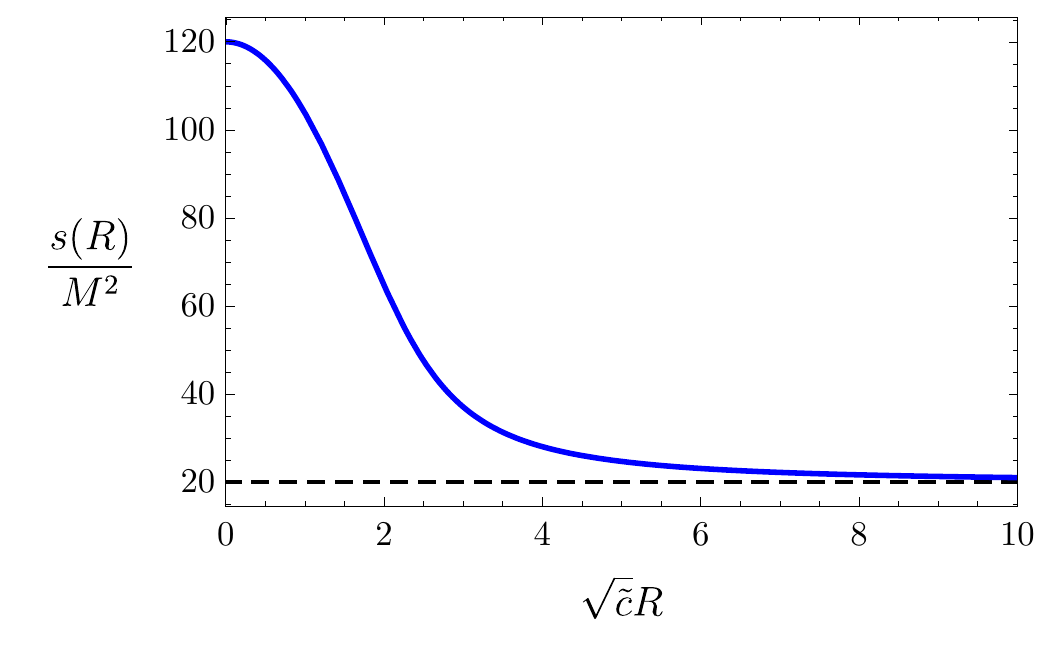}
		\caption{\(\ct > 0,~\dfrac{N}{M} = 10\)}
	\end{subfigure}
	\\[1em]
	\begin{subfigure}{0.5\textwidth}
		\includegraphics[width=\textwidth]{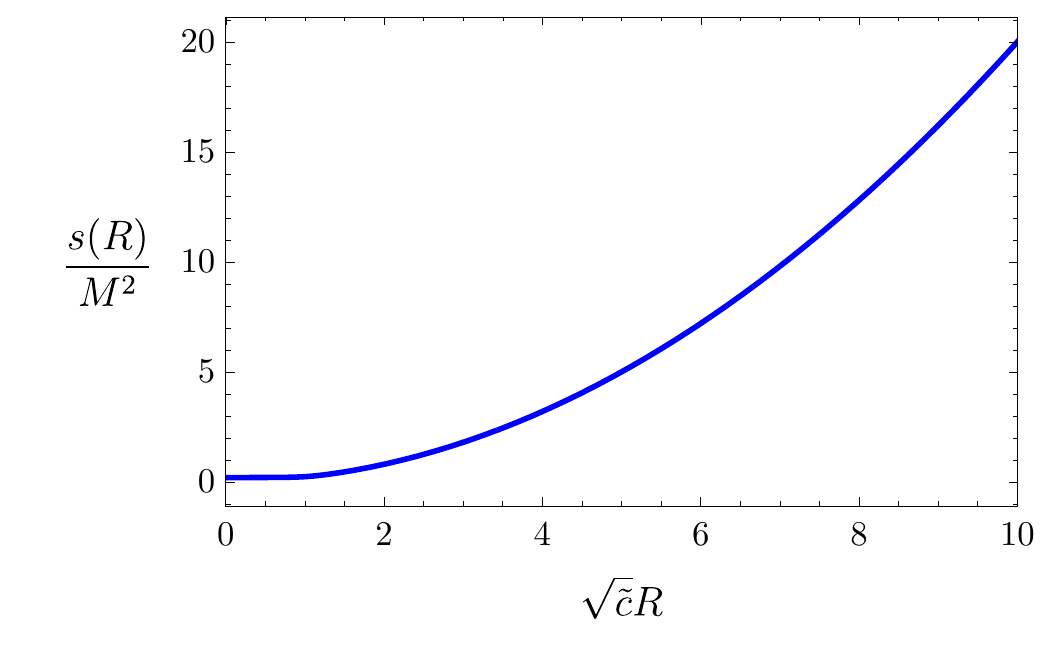}
		\caption{\(\ct < 0,~\dfrac{N}{M} = \dfrac{1}{10}\)}
	\end{subfigure}
	\begin{subfigure}{0.5\textwidth}
		\includegraphics[width=\textwidth]{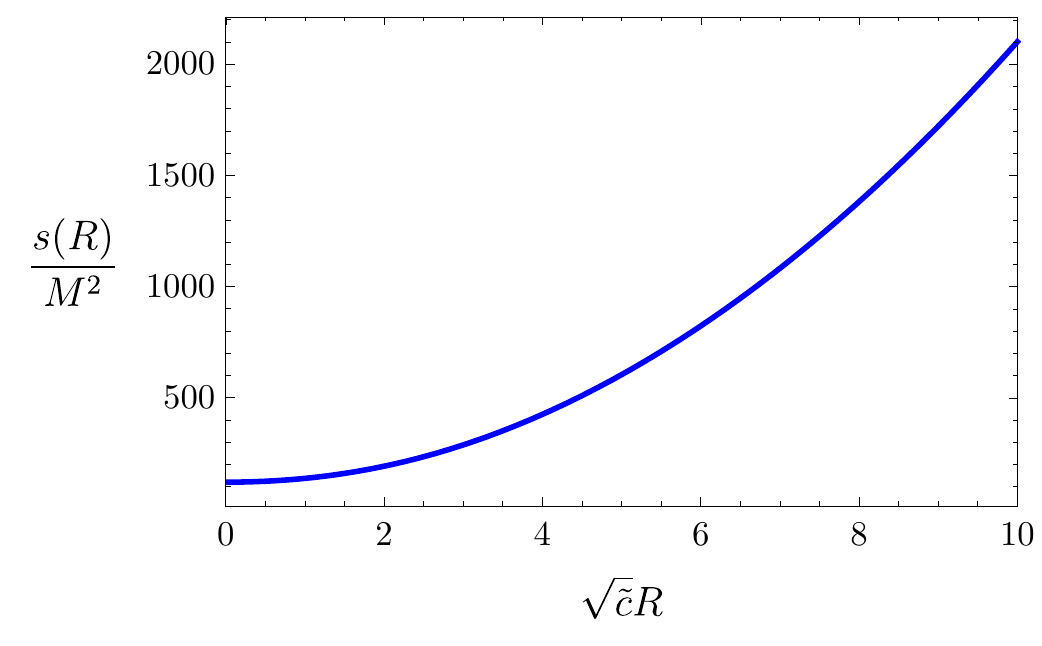}
		\caption{\(\ct < 0,~\dfrac{N}{M} = 10\)}
	\end{subfigure}
	\caption{
			The derivative with repsect to \(\log R\) of the contribution to the entanglement wedge on-shell action of the symmetric flow (top row) and funnel (bottom row) solutions, for sample values of \(N/M\). For \(\ct < 0\), corresponding to a flow from a symmetric representation Wilson surface in the UV to a bundle of M2-branes in the IR, the derivative interpolates monotonically between the values at the fixed points, given in \eqref{eq:symmetric_flow_s}. The horizontal dashed line shows the value at the IR fixed point. For \(\ct > 0\), corresponding to a funnel solution, the coefficient of the logarithm in the on-shell action increases monotonically without bound.
		}
	\label{fig:symmetric_flow_action}
\end{figure}

Computing the logarithmic derivative with respect to the radius of the entangling region, we obtain the UV finite quantity \(s(R)\), which we write as
\begin{equation} \label{eq:symmetric_flow_s}
	s(R)
	\equiv R \frac{\diff S_\mathcal{W}^\star}{\diff R} 
	= M N \le(
		1 + \frac{N}{2M} - \ct R^2
		+ \sqrt{\le(1 + \frac{N}{2M} - \ct R^2 \ri)^2 + 4 \ct R^2}
	\ri). 
\end{equation}
We plot the form of this function for sample values of \(N/M\) in figure~\ref{fig:symmetric_flow_action}. It is bounded from below by \(2 M N\),\footnote{
	To see this, rewrite the function as \(s(R)
	= M N \le[
		2
		+ \sqrt{\le(1 - \frac{N}{2M} + \ct R^2 \ri)^2 + \frac{2N}{M}} - \le(1 - \frac{N}{2M} + \ct R^2 \ri)
	\ri]\).}
and it is straightforward to show that it monotonically decreases with \(R\) for the symmetric flow solution (\(\ct > 0\)), and monotonically increases for the funnel solution (\(\ct < 0\)). To do so, we note that \(R\) appears only in the dimensionless combination \(\ct R^2\), and
\begin{equation}
	\frac{\diff s}{\diff (\ct R^2)} = \frac{2 M N - s}{M N \sqrt{\le(1 - \frac{N}{2M} + \ct R^2 \ri)^2 + \frac{2N}{M}}} < 0.
\end{equation}

For small \(R\), we find that \(s\) is given by the value for a symmetric representation Wilson surface,
\begin{equation}
	s(R=0) = 2 N \le(
		M + \frac{N}{2}
	\ri).
\end{equation}
The behavior at large \(R\) depends on the sign of \(\ct\),
\begin{equation}
	s(R \to \infty) \sim \begin{cases}
		2 M N, \quad & \ct > 0,
		\\
		2 M N |\ct| R^2, & \ct < 0.
	\end{cases}
\end{equation}
In particular, the large \(R\) limit for \(\ct > 0\) is \(N\) times the value for a single M2-brane. For both the antisymmetric and symmetric flow solutions, the entanglement wedge on-shell action provides a quantity which decreases monotonically under RG flows.

%%%%%%%%%%%%%%%%%%%%%%%%%%%%%%%%%%%%%%%%%%%%%%%%%%
%%%%%%%%%%%%%%%%%%%%%%%%%%%%%%%%%%%%%%%%%%%%%%%%%%
\section{Discussion}
\label{sec:discussion}
%%%%%%%%%%%%%%%%%%%%%%%%%%%%%%%%%%%%%%%%%%%%%%%%%%
%%%%%%%%%%%%%%%%%%%%%%%%%%%%%%%%%%%%%%%%%%%%%%%%%%

We have computed the contribution to entanglement entropy from a number of defects in \(\mathcal{N}=(2,0)\) SCFT, holographically dual to probe M-theory branes, for a spherical entangling region centred on the defect. Some of these defects were Wilson surfaces, and for these the entanglement entropy reproduces the probe limit of the results of \cite{us_entanglement}.

The contribution of a two-dimensional conformal defect to the entanglement entropy of a spherical subregion takes the same form as the entanglement entropy of a single interval in a two-dimensional CFT, and in particular is logarithmically divergent in the UV. It is thus natural to identify the coefficient \(b\) of the logarithm as a central charge for the defect, and moreover the \(b\)-function~\eqref{eq:b_function} provides a natural quantity that interpolates between the central charges of the fixed points of a defect RG flow.

The M5-brane embeddings we have studied show that \(b\) is not necessarily monotonic along RG flows, and in particular can be larger in the IR than in the UV. This suggests that the central charge as defined from the entanglement entropy may not provide a measure of the number of massless degrees of freedom on the defect.

An alternative quantity, the on-shell action inside the entanglement wedge inside, decreases monotonically along the flows we study, as well as similar flows involving D-branes~\cite{Kumar:2017vjv}. This provides another candidate \(C\)-function. It would be interesting to test whether it is monotonic in other holographic examples of RG flows, and to understand what this quantity corresponds to in the dual field theory.

There are several interesting directions for future work. We have studied planar defects, and a natural generalization would be to study the entanglement entropy of defects with more complicated geometries. One example is the spherical Wilson surface, which may be obtained from the planar surface by a conformal transformation \cite{Chen:2007ir}. One could also study different geometries for the entangling region, or defects in holographic examples of six-dimensional SCFTs with \(\mathcal{N} = (1,0)\) supersymmetry \cite{Gaiotto:2014lca}.

In a 2D CFT, single-interval entanglement entropy is determined by the central charge, which appears in many other quantities including the Weyl anomaly and the thermodynamic entropy. For a higher dimensional CFT with a two dimensional defect or boundary, it is currently unknown whether the central charge \(b\) that we have defined using entanglement entropy appears in any other quantities.

%%%%%%%%%%%%%%%%%%%%%%%%%%%%%%%%%%%%%%%%%%%%%%%%%%
%%%%%%%%%%%%%%%%%%%%%%%%%%%%%%%%%%%%%%%%%%%%%%%%%%
\section*{Acknowledgements}
%%%%%%%%%%%%%%%%%%%%%%%%%%%%%%%%%%%%%%%%%%%%%%%%%%
%%%%%%%%%%%%%%%%%%%%%%%%%%%%%%%%%%%%%%%%%%%%%%%%%%

I have benefitted from interesting discussions with Matthew Buican, Nadav Drukker, Matti Jarvinen, Carlos N\'u\~nez, and Daniel Thompson. I would especially like to thank John~Estes, Darya~Krym, S. Prem Kumar, Andy~O'Bannon, and Brandon~Robinson for frequent and helpful discussions, and for comments on a draft of the manuscript. This work was supported by an STFC studentship through Consolidated Grant ST/L000296/1.

%%%%%%%%%%%%%%%%%%%%%%%%%%%%%%%%%%%%%%%%%%%%%%%%%%
%%%%%%%%%%%%%%%%%%%%%%%%%%%%%%%%%%%%%%%%%%%%%%%%%%
\appendix
%%%%%%%%%%%%%%%%%%%%%%%%%%%%%%%%%%%%%%%%%%%%%%%%%%
%%%%%%%%%%%%%%%%%%%%%%%%%%%%%%%%%%%%%%%%%%%%%%%%%%

%%%%%%%%%%%%%%%%%%%%%%%%%%%%%%%%%%%%%%%%%%%%%%%%%%
%%%%%%%%%%%%%%%%%%%%%%%%%%%%%%%%%%%%%%%%%%%%%%%%%%
\section{Details of entanglement entropy calculations}
%%%%%%%%%%%%%%%%%%%%%%%%%%%%%%%%%%%%%%%%%%%%%%%%%%
%%%%%%%%%%%%%%%%%%%%%%%%%%%%%%%%%%%%%%%%%%%%%%%%%%

%%%%%%%%%%%%%%%%%%%%%%%%%%%%%%%%%%%%%%%%%%%%%%%%%%
%%%%%%%%%%%%%%%%%%%%%%%%%%%%%%%%%%%%%%%%%%%%%%%%%%
\subsection{Entanglement entropy of the antisymmetric flow solution}
\label{app:antisymmetric_flow}
%%%%%%%%%%%%%%%%%%%%%%%%%%%%%%%%%%%%%%%%%%%%%%%%%%
%%%%%%%%%%%%%%%%%%%%%%%%%%%%%%%%%%%%%%%%%%%%%%%%%%

Differentiating the off-shell action \eqref{eq:antisymmetric_flow_hyperbolic_action} with respect to the inverse temperature \(\b\), and making use of \eqref{eq:hyperbolic_ee_off_shell} we find that the entanglement entropy may be written as
\begin{subequations}
\begin{equation}
	\see^{(1)} = \see^\mathrm{horizon} + \see^\mathrm{bulk},
\end{equation}
with horizon and bulk contributions given respectively by 
\begin{align}
	\see^\mathrm{horizon} &= \frac{1}{5} \int_0^{2\pi} \diff \t \int_0^{u_c} \diff u \le. \mathcal{L} \ri|_{v=v_H=1},
	\\
	\see^\mathrm{bulk} &= - \frac{1}{5} \int_0^{2\pi} \diff \t \int_1^\infty \diff v \int_0^{u_c} \diff u \lim_{v_H \to 1} \p_{v_H} \mathcal{L},
\end{align}
\end{subequations}
where \(\mathcal{L}\) is given by \eqref{eq:antisymmetric_flow_hyperbolic_lagrangian}, and \(\q\) is to be taken on-shell after the differentiation with respect to \(v_H\) is performed.

By performing the coordinate transformation back to flat slicing using the inverse of the map \eqref{eq:map_to_hyperbolic}, the combination of derivatives appearing in \(\mathcal{L}\) may be written as (for \(v_H = 1\))
\begin{equation}
	4 + \frac{1}{f(v)} (\p_\t \q)^2 + f(v) (\p_v \q)^2 + \frac{1}{v^2} (\p_u \q)^2 = 4 + z^2 \q'(z)^2 = 4 \frac{D(\q)^2 + \sin^6 \q}{\le(
		D(\q) \cos \q + \sin^4 \q
	\ri)^2},	
\end{equation}
where we have made use of the BPS condition \eqref{eq:flow_bps}. This simplifies the integrands slightly, so that
\begin{subequations}
\begin{align}
	\see^\mathrm{horizon} &= \frac{M^2}{5} \int_0^{u_c} \diff u 
	\frac{
		D(\q)^2 + \sin^6 \q
	}{
		D(\q) \cos \q + \sin^4 \q
	},
	\\
	\see^\mathrm{brane} &= \frac{M^2}{20\pi} \int_0^{2\pi} \diff \t \int_1^\infty \diff v \int_0^{u_c} \diff u \frac{1}{v^3}\le[
		(\p_v \q)^2 - \frac{(\p_\t \q)^2}{(v^2 - 1)^2}
	\ri] \le(
		D(\q) \cos \q + \sin^4 \q
	\ri).
\end{align}
\end{subequations}
We now change integration variables back from \((\t,v,u)\) to the flat slicing coordinates \((x^0,x,z)\). Once more making use of the BPS condition \eqref{eq:flow_bps}, we find that the integrals may be written as
\begin{subequations}
	\label{eq:antisymmetric_flow_ee_intermediate}
	\begin{align}
	\see^\mathrm{horizon} &= \frac{2M^2}{5} \int_\e^R \diff z \frac{R}{z \sqrt{R^2 - z^2}}
	\frac{
		D(\q)^2 + \sin^6 \q
	}{
		D(\q) \cos \q + \sin^4 \q
	},
	\\
	\see^\mathrm{brane} &= \frac{4 M^2}{5\pi} \int \diff x^0 \diff x \diff z \frac{1}{z^5 v^2 (v^2 - 1)}\le[
		(v^2 - 1)^2 (\p_v z)^2 - (\p_\t z)^2
	\ri]
	\nonumber \\ &\hspace{6cm}
	\times \frac{
		\le( D(\q) \sin\q - \cos\q \sin^3\q \ri)^2
	}{
		D(\q) \cos\q + \sin^4 \q
	}.
	\label{eq:antisymmetric_flow_ee_intermediate_bulk}
	\end{align}
\end{subequations}
In Euclidean signature, the image of the inverse of the map \eqref{eq:map_to_hyperbolic} is all of local \(AdS\), rather than the region \(x_0^2 + x^2 + r^2 + z^2 \leq R^2\) as is the case in Lorentzian signature. The integration region in the bulk contribution \eqref{eq:antisymmetric_flow_ee_intermediate_bulk} is therefore the entirety of local \(AdS\), with the cutoff region at~\(z<\e\) excised.

The derivatives of \(z\) with respect to hyperbolic slicing coordinates may be written as 
\begin{align}
	\p_\t z = \frac{x^0 z}{R},
	\quad
	(v^2 - 1)\p_v z = \frac{1}{2 R z} \frac{
		z^4 - \le[x_0^2 + (R - x)^2\ri]\le[x_0^2 + (R + x)^2\ri]
	}{
		\sqrt{(R^2 - x_0^2 - x^2 - z^2) + 4 R^2 (x_0^2 + z^2)}
	}.
\end{align}
Substituting these into the bulk integral in \eqref{eq:antisymmetric_flow_ee_intermediate} yields the final form \eqref{eq:antisymmetric_flow_ee}.

%%%%%%%%%%%%%%%%%%%%%%%%%%%%%%%%%%%%%%%%%%%%%%%%%%
%%%%%%%%%%%%%%%%%%%%%%%%%%%%%%%%%%%%%%%%%%%%%%%%%%
\subsection{Entanglement entropy for M5-branes wrapping \(S^3 \subset AdS_7\)}
\label{app:spike_entanglement}
%%%%%%%%%%%%%%%%%%%%%%%%%%%%%%%%%%%%%%%%%%%%%%%%%%
%%%%%%%%%%%%%%%%%%%%%%%%%%%%%%%%%%%%%%%%%%%%%%%%%%

In the hyperbolic slicing of \(AdS\) \eqref{eq:hyperbolic_slicing}, parameterising the M5-brane by \((\t,v,u,\f_1,\f_2,\f_3)\) the solution~\eqref{eq:spike_solution} becomes
\begin{equation} \label{eq:spike_solution_hyperbolic}
    \sin\f_0 \equiv \Phi= \frac{\kappa}{ v \sinh u \sqrt{1 + \ct R^2 \Omega^{-2} }},
    \quad
    \Omega = v \cosh u + \sqrt{v^2 - 1} \cos \t.
\end{equation}

We again split the entanglement entropy into horizon and bulk contributions, \(\see^{(1)} = \see^\mathrm{horizon} + \see^\mathrm{bulk}\), with
\begin{subequations} \label{eq:spike_ee_split}
\begin{align}
	\see^\mathrm{horizon} &= \frac{1}{5} \int_0^{2\pi} \diff \t \int_{u_\mathrm{min}}^{u_c} \diff u \le. \mathcal{L} \ri|_{v=v_H=1},
	\\
	\see^\mathrm{bulk} &= - \frac{1}{5} \int_0^{2\pi} \diff \t \int_1^\infty \diff v \int_{u_\mathrm{min}}^{u_c} \diff u \lim_{v_H \to 1} \p_{v_H} \mathcal{L},
\end{align}
\end{subequations}
where \(\mathcal{L}\) is given by \eqref{eq:symmetric_off_shell_lagrangian}, \(u_\mathrm{min}\) is determined by the requirement that \(\F \leq 1\), and \(\F\) is to be taken on-shell only after the differentiation with respect to \(v_H\) is performed.

The solution~\eqref{eq:spike_solution_hyperbolic} satisfies,
\begin{equation}
    1 - \F^2  + v^2 \sinh^2 u \hat{g}^{ab} \p_a \F \p_b \F
    =
    1 + \frac{\F^6 v^6 \sinh^6 u}{\kappa^4},
\end{equation}
which simplifies the integrands appearing in~\eqref{eq:spike_entanglement_integral}
\begin{subequations}
\begin{align}
    \see^\mathrm{horizon} &= \frac{8 M N}{5} \int_{u_\mathrm{min}}^\infty \diff u
    \le. \frac{1}{\sqrt{1 - \F^2} }
    \le(
        1 + \frac{\F^6 v^6 \sinh^6 u}{\kappa^4}
    \ri) \ri|_{v=1},
    \\
    \see^\mathrm{bulk} &= \frac{8 M^2}{5 \pi} \int \diff \t \diff v \diff u \frac{1}{\sqrt{1 - \F^2}} \le\{
        \frac{\kappa^2 \sinh^2 u}{v} \le[
            (\p_v \F)^2 - \frac{(\p_\t \F)^2}{(v^2 - 1)^2}
        \ri]
        + 6 \sinh^4 u \F^3 \p_v \F
    \ri\}.
\end{align}
\end{subequations}

As for the antisymmetric flow solution, it will be convenient to perform a coordinate transformation back to flat slicing. Rather than use the chain rule, to transform the derivatives appearing in the bulk integral, we note that the solution~\eqref{eq:spike_solution_hyperbolic} satisfies 
\begin{subequations}
\begin{align}
    \p_\t \F &= - \frac{\kappa \sqrt{v^2 - 1} \sin \t}{\sqrt{\ct} R v \sinh u} \le[
        1 - \le(
            \frac{\F v \sinh u}{\kappa}
        \ri)^2
    \ri]^{3/2},
    \\
    \p_v \F &= \frac{\kappa \le(
        \sqrt{v^2 - 1} \cosh u + v \cos \t
    \ri)}{
        \sqrt{\ct} R v \sqrt{v^2 - 1} \sinh u
    } \le[
        1 - \le(
            \frac{\F v \sinh u}{\kappa}
        \ri)^2
    \ri]^{3/2} - \frac{\F}{v}.
\end{align}
\end{subequations}
In flat slicing, the solution becomes
\begin{equation} \label{eq:phi_flat_slicing}
    \F = \frac{\kappa z}{
        \sqrt{(1 + \ct z^2)x^2 + \kappa^2 z^2}.
    }
\end{equation}
Performing the inverse of the transformation~\eqref{eq:map_to_hyperbolic} and substituting the solution~\eqref{eq:phi_flat_slicing}, we find that the derivatives become
\begin{subequations}
    \begin{align}
        \p_\t \F &= - \frac{ \kappa \ct x^0 z^3 }{
            R (1 + \ct z^2) \sqrt{(1 + \ct z^2) x^2 + \kappa^2 z^2}
        },
        \\
        \p_v \F &= -\frac{\kappa z}{v} \frac{
            \le( R^2 - x_0^2 - x^2 - r^2 \ri)^2
            + 4 R^2 x_0^2 - z^4
            - \frac{2 \kappa^2 z^4}{r^2} \le(
                R^2 - x_0^2 - x^2 - r^2 - z^2
            \ri)
        }{
            \sqrt{(1 + \ct z^2)x^2 + \kappa^2 z^2}
            ( 1 + \ct z^2 )
            \le[
                4 R^2 x_0^2 + (R^2 - x_0^2 - x^2 - r^2 - z^2)^2
            \ri]
        }.
    \end{align}
\end{subequations}
Plugging these into the integrals in~\eqref{eq:spike_ee_split} and performing the coordinate transformation \((\t,v,u) \to (x^0,x,z)\), we obtain~\eqref{eq:spike_entanglement_integral}.

%%%%%%%%%%%%%%%%%%%%%%%%%%%%%%%%%%%%%%%%%%%%%%%%%%
%%%%%%%%%%%%%%%%%%%%%%%%%%%%%%%%%%%%%%%%%%%%%%%%%%
\bibliographystyle{JHEP}
\bibliography{m5_ref}
%%%%%%%%%%%%%%%%%%%%%%%%%%%%%%%%%%%%%%%%%%%%%%%%%%
%%%%%%%%%%%%%%%%%%%%%%%%%%%%%%%%%%%%%%%%%%%%%%%%%%

\end{document}